\documentclass[a4paper,11pt]{article}
\pdfoutput=1 

\usepackage{jheppub} 
\makeatletter
\DeclareRobustCommand*{\bfseries}{%
  \not@math@alphabet\bfseries\mathbf
  \fontseries\bfdefault\selectfont
  \boldmath
}
\makeatother

\usepackage{calc}
\usepackage{rotating}
\usepackage[english]{babel}
\usepackage{graphicx}
\usepackage{subfig}
\usepackage{float}
\usepackage{amsmath}
\usepackage{amsmath,bm}
\usepackage{amssymb}
\usepackage{amsthm}
\usepackage{latexsym}
\usepackage{color}
\usepackage{mciteplus}
\usepackage{fancyhdr}
\usepackage{multicol}
\usepackage{booktabs}
\usepackage{colortbl}
\usepackage{hyperref}
\usepackage{diagbox}
\usepackage[inline]{enumitem}
\usepackage{ulem}
\usepackage{longtable}
\usepackage{caption}
\newcommand{\newc}{\newcommand*}

\long\def\begincomment#1\endcomment{%
        \begingroup\sf\baselineskip12pt#1\endgroup}

\newc{\etal}{\textrm{et al.}} 
\newc{\eg}{\textrm{e.g.}} 
\newc{\ie}{\textrm{i.e.}}
\newc{\etc}{\textrm{etc}}
\newc\vs{\textrm{vs.}}
\newc{\cl}{\rm {C.L.}}

\newc{\ev}{\ensuremath{\,\mathrm{eV}}}
\newc{\kev}{\ensuremath{\,\mathrm{keV}}}
\newc{\mev}{\ensuremath{\,\mathrm{MeV}}}
\newc{\gev}{\ensuremath{\,\mathrm{GeV}}}
\newc{\tev}{\ensuremath{\,\mathrm{TeV}}}
\newc{\MeV}{\mev} 
\newc{\TeV}{\tev}
\newc{\invpb}{\ensuremath{/\text{pb}}}
\newc{\invfb}{\ensuremath{/\text{fb}}}
\newc\nb{\ensuremath{\,\mathrm{nb}}} \newc\pb{\ensuremath{\,\mathrm{pb}}} \newc\fb{\ensuremath{\,\mathrm{fb}}}
\newc\pc{\ensuremath{\,\mathrm{pc}}}
\newc\kpc{\ensuremath{\,\mathrm{kpc}}}
\newc\mpc{\ensuremath{\,\mathrm{Mpc}}}
\newc\ps{\ensuremath{\,\mathrm{ps}}} 
\newc\cmeter{\ensuremath{\,\mathrm{cm}}} 
\newc\meter{\ensuremath{\,\mathrm{m}}} 
\newc\kmeter{\ensuremath{\,\mathrm{km}}}
\newc\second{\ensuremath{\,\mathrm{s}}}
\newc\msecond{\ensuremath{\,\mathrm{ms}}}
\newc\nsecond{\ensuremath{\,\mathrm{ns}}}
\newc\psecond{\ensuremath{\,\mathrm{ps}}}

\newc{\chisqmin}{\ensuremath{\chi^2_{\mathrm{min}}}}
\newc{\Delchisq}{\ensuremath{\Delta\chi^2}}
\newc{\chisq}{\ensuremath{\chi^2}}
\newc{\like}{\ensuremath{\mathcal{L}}}

\newc\lsim{\ensuremath{\mathrel{\rlap{\lower4pt\hbox{\hskip1pt$\sim$}}\raise1pt\hbox{$<$}}}}
\newc\gsim{\ensuremath{\mathrel{\rlap{\lower4pt\hbox{\hskip1pt$\sim$}}\raise1pt\hbox{$>$}}}}
\newc{\VEV}[1]{\ensuremath{\langle #1 \rangle}}
\newc{\dl}{\ensuremath{\stackrel{\leftarrow}{D}}}
\newc{\dr}{\ensuremath{\stackrel{\rightarrow}{D}}}

\newc{\bcenter}{\begin{center}}    \newc{\ecenter}{\end{center}}
\newc{\bfl}{\begin{flushleft}}    \newc{\efl}{\end{flushleft}}
\newc{\bfr}{\begin{flushright}}    \newc{\efr}{\end{flushright}}

\newc{\bi}{\begin{itemize}}
\newc{\ei}{\end{itemize}}
\newc{\bed}{\begin{description}}
\newc{\eed}{\end{description}}
\newc{\ben}{\begin{enumerate}}
\newc{\een}{\end{enumerate}}

\newc{\be}{\begin{equation}}
\newc{\ee}{\end{equation}}
\newc{\bea}{\begin{eqnarray}}
\newc{\eea}{\end{eqnarray}}
\newc{\bfle}{\begin{flalign}}
\newc{\efle}{\end{flalign}}
\newc{\ra}{\rightarrow}

\newc{\alphas}{\ensuremath{\alpha_s}}
\newc{\alphatwo}{\ensuremath{\alpha_2}}
\newc{\alphaone}{\ensuremath{\alpha_1}}
\newc{\alphai}[1]{\ensuremath{\alpha_{#1}}}
\newc{\alphaem}{\ensuremath{\alpha_{\mathrm{em}}}}
\newc{\alphaeff}{\ensuremath{\alpha_{\mathrm{eff}}}}
\newc{\sineff}{\ensuremath{\sin \theta_{\mathrm{eff}}}}
\newc{\sinsqeff}{\ensuremath{\sin^2 \theta_{\mathrm{eff}}}}
\newc{\dalphahad}{\ensuremath{\Delta \alpha_{\mathrm{had}}}}
\newc{\yt}{\ensuremath{h_t}} \newc{\yb}{\ensuremath{h_b}} \newc{\ytau}{\ensuremath{h_{\tau}}}
\newc\mz{\ensuremath{M_Z}} 
\newc\mw{\ensuremath{m_W}}
\newc\mZ{\mz}        \newc\mW{\mw}
\newc\mhsm{\ensuremath{ m_{H_{\mathrm{SM}}}}}
\newc{\mtop}{\ensuremath{ m_t}}               \newc{\mtpole}{\ensuremath{ M_t}}
\newc{\mbottom}{\ensuremath{ m_b}} 
\newc{\mtau}{\ensuremath{ m_{\tau}}}
\newc{\mt}{\mtpole}
\newc{\mb}{\mbottom} 
\newc{\rtwogg}{\ensuremath{R_{h_2}(\gamma\gamma)}}
\newc{\rtwozz}{\ensuremath{R_{h_2}(ZZ)}}
\newc{\ronegg}{\ensuremath{R_{h_1}(\gamma\gamma)}}
\newc{\ronezz}{\ensuremath{R_{h_1}(ZZ)}}
\newc{\rsiggg}{\ensuremath{R_{h_\textrm{sig}}(\gamma\gamma)}}
\newc{\rsigzz}{\ensuremath{R_{h_\textrm{sig}}(ZZ)}}
\newc{\llbar}{\ensuremath{\ell\bar{\ell}}}
\newc{\tauptaum}{\ensuremath{ \tau^+\tau^-}}
\newc{\qqbar}{\ensuremath{ q\bar{q}}} \newc{\ppbar}{\ensuremath{ p\bar{p}}}
\newc{\bbbar}{\ensuremath{ b\bar{b}}} \newc{\ttbar}{\ensuremath{ t\bar{t}}}
\newc{\ffbar}{\ensuremath{ f\bar{f}}} \newc{\tautaubar}{\ensuremath{ \tau\bar{\tau}}}

\newc{\mchi}{\ensuremath{m_\neutone}}
\newc{\squark}{\ensuremath{\tilde{q}}}
\newc{\slepton}{\ensuremath{\tilde{l}}}
\newc{\gluino}{\ensuremath{\tilde{g}}} 
\newc{\mgluino}{\ensuremath{{m_{\gluino}}}}
\newc{\wino}{\ensuremath{\tilde{W}}} 
\newc{\mwino}{\ensuremath{{m_{\wino}}}}
\newc{\tone}{\ensuremath{{\tilde{t}_1}}}
\newc{\Hone}{\ensuremath{{\tilde{H}_{1}}}}
\newc{\Htwo}{\ensuremath{{\tilde{H}_{2}}}}
\newc{\Hhtwo}{\ensuremath{{H_{2}}}}
\newc{\qli}{\ensuremath{{\tilde{Q}_{i}}}}
\newc{\uri}{\ensuremath{{\tilde{u}_{i}}}}
\newc{\dri}{\ensuremath{{\tilde{d}_{i}}}}
\newc{\lli}{\ensuremath{{\tilde{L}_{i}}}}
\newc{\eri}{\ensuremath{{\tilde{e}_{i}}}}

\newc{\sthw}{\ensuremath{ \sin\theta_W}}              \newc{\cthw}{\ensuremath{\cos\theta_W}}
\newc{\tanthw}{\ensuremath{ \tan\theta_W}}              \newc{\cotthw}{\ensuremath{\cot\theta_W}}
\newc{\ssqthw}{\ensuremath{\sin^2 \theta_W}}
\newc{\msbar}{\ensuremath{\overline{MS}}} \newc{\drbar}{\ensuremath{\overline{DR}}}
\newc{\mtmtsmmsbar}{\ensuremath{ m_t(m_t)^{\msbar}_{{\mathrm{SM}}}}}
\newc{\mtmtsmdrbar}{\ensuremath{ m_t(m_t)^{\drbar}_{{\mathrm{SM}}}}}
\newc{\mtmtmssmdrbar}{\ensuremath{ m_t(m_t)^{\drbar}_{{\mathrm{SUSY}}}}}
\newc{\mbmbmsbar}{\ensuremath{ m_b(m_b)^{\msbar} }}
\newc{\mbmbsmmsbar}{\ensuremath{ m_b(m_b)^{\msbar}_{{\mathrm{SM}}}}}
\newc{\mbmzsmmsbar}{\ensuremath{ m_b(\mz)^{\msbar}_{{\mathrm{SM}}}}}
\newc{\mbmzsmdrbar}{\ensuremath{ m_b(\mz)^{\drbar}_{{\mathrm{SM}}}}}
\newc{\mbmzmssmdrbar}{\ensuremath{ m_b(\mz)^{\drbar}_{{\mathrm{SUSY}}}}}
\newc{\mtaumzsmmsbar}{\ensuremath{ m_{\tau}(\mz)^{\msbar}_{{\mathrm{SM}}}}}
\newc{\mtaumzsmdrbar}{\ensuremath{ m_{\tau}(\mz)^{\drbar}_{{\mathrm{SM}}}}}
\newc{\mtaumzmssmdrbar}{\ensuremath{ m_{\tau}(\mz)^{\drbar}_{{\mathrm{SUSY}}}}}
\newc{\alphasmzms}{\ensuremath{\alpha_s(M_Z)^{\overline{MS}}}}
\newc{\alphaimzms}[1]{\ensuremath{\alpha_{#1}(M_Z)^{\overline{MS}}}}
\newc{\alphaemmz}{\ensuremath{\alpha_{\mathrm{em}}(M_Z)^{\overline{MS}}}}

\newc{\mzero}{\ensuremath{{m_0}}}
\newc{\mhalf}{\ensuremath{ m_{1/2}}}
\newc{\tanb}{\ensuremath{\tan\beta}}
\newc{\azero}{\ensuremath{ A_0}}
\newc{\signmu}{\ensuremath{\rm{sgn}\,\mu}}
\newc{\atau}{\ensuremath{{A_{\tau}}}}
\newc{\mueff}{\ensuremath{\mu_{\rm{eff}}}}
\newc{\lam}{\ensuremath{{\lambda}}}
\newc{\kap}{\ensuremath{{\kappa}}}
\newc{\alam}{\ensuremath{{A_{\lambda}}}}
\newc{\akap}{\ensuremath{{A_{\kappa}}}}
\newc{\hs}{\ensuremath{ H_s}}      
\newc{\mhs}{\ensuremath{ m_{H_s}}} 
\newc{\mgut}{\ensuremath{ M_{\rm GUT}}}
\newc{\mvl}{\ensuremath{ M_{\rm VL}}}
\newc{\gut}{\ensuremath{{\rm GUT}}}
\newc{\mplanck}{\ensuremath{ M_{\rm P}}}      \newc{\mpl}{\ensuremath{ M_{\rm Pl}}}
\newc{\msusy}{\ensuremath{ M_{\rm SUSY}}}      \newc{\ms}{\ensuremath{ M_{\rm S}}}
 \newc{\hu}{\ensuremath{ H_u}}       \newc{\hd}{\ensuremath{ H_d}}
 \newc{\mhu}{\ensuremath{ m_{H_u}}}       \newc{\mhd}{\ensuremath{ m_{H_d}}}
 \newc{\mhuew}{\ensuremath{ m^{\ast}_{H_u}}}       \newc{\mhdew}{\ensuremath{ m^{\ast}_{H_d}}}
 \newc{\mhuewsq}{\ensuremath{ m^{\ast\, 2}_{H_u}}}       \newc{\mhdewsq}{\ensuremath{ m^{\ast\, 2}_{H_d}}}
 \newc{\mhl}{\ensuremath{m_\hl}} 
 \newc{\mhone}{\ensuremath{m_{h_1}}} 
 \newc{\mhtwo}{\ensuremath{m_{h_2}}} 
 \newc{\mhi}{\ensuremath{m_{\tilde{h}}}} 
 \newc{\mul}{\ensuremath{m_{\tilde{u}_L}}} 
 \newc{\mtone}{\ensuremath{m_{\tilde{t}_1}}} 
 \newc{\ma}{\ensuremath{m_A}} 
 \newc{\mH}{\ensuremath{m_H}} 
 \newc{\maone}{\ensuremath{m_{a_1}}} 
 \newc{\matwo}{\ensuremath{m_{a_2}}}
 \newc{\hone}{\ensuremath{h_1}}
 \newc{\htwo}{\ensuremath{h_2}}
 \newc{\aone}{\ensuremath{a_1}}
 \newc{\atwo}{\ensuremath{a_2}}
 \newc{\mqthree}{\ensuremath{m_{\tilde{Q}_3}^2}}
 \newc{\muthree}{\ensuremath{m_{\tilde{u}_3}^2}}
 \newc{\mqli}{\ensuremath{m_{\tilde{Q}_{i}}}}
 \newc{\muri}{\ensuremath{m_{\tilde{u}_{i}}}}
 \newc{\mdri}{\ensuremath{m_{\tilde{d}_{i}}}}
 \newc{\mlli}{\ensuremath{m_{\tilde{L}_{i}}}}
 \newc{\meri}{\ensuremath{m_{\tilde{e}_{i}}}}
 \newc{\ts}{\ensuremath{T_{SUSY}}}

\newc{\sigsip}{\ensuremath{\sigma^{\rm SI}_{p}}}	\newc{\sigsin}{\ensuremath{\sigma^{\rm SI}_{n}}}
\newc{\sigsdp}{\ensuremath{\sigma^{\rm SD}_{p}}}	\newc{\sigsdn}{\ensuremath{\sigma^{\rm SD}_{n}}}
\newc{\sigsi}{\ensuremath{\sigma^{\rm SI}}}	\newc{\sigsd}{\ensuremath{\sigma^{\rm SD}}}
\newc{\abund}{\ensuremath{ \Omega h^2}}
\newc{\omegadm}{\ensuremath{ \Omega_{{\rm DM}}}}     \newc{\abunddm}{\ensuremath{ \Omega_{{\rm DM}} h^2}} 
\newc{\omegam}{\ensuremath{ \Omega_{{\rm m}}}}       \newc{\abundm}{\ensuremath{ \Omega_{{\rm m}} h^2}}
\newc{\omegab}{\ensuremath{ \Omega_{{\rm b}}}}	\newc{\abundb}{\ensuremath{ \Omega_{{\rm b}} h^2}}
\newc{\omegatot}{\ensuremath{ \Omega_{{\rm TOT}}}}
\newc{\omegacdm}{\ensuremath{ \Omega_{{\rm CDM}}}}   \newc{\abundcdm}{\ensuremath{ \Omega_{{\rm CDM}} h^2}}
\newc{\omegalambda}{\ensuremath{ \Omega_{\Lambda}}} \newc{\abundlambda}{\ensuremath{ \Omega_{\Lambda} h^2}}
\newc{\omegarad}{\ensuremath{ \Omega_{{\rm rad}}}}  \newc{\abundrad}{\ensuremath{ \Omega_{{\rm rad}} h^2}}
\newc{\rhocrit}{\ensuremath{ \rho_{\rm crit}}}
\newc{\rhochi}{\ensuremath{ \rho_{\chi}}}
\newc{\abunchi}{\ensuremath{\Omega_\chi h^2}}
\newc{\abundlsp}{\ensuremath{\Omega_{\rm LSP}h^2}}

\newc{\amu}{\ensuremath{ a_{\mu}}}        \newc{\amususy}{\ensuremath{ a_{\mu}^{\mathrm{SUSY}}}}
\newc{\amuexpt}{\ensuremath{ a_{\mu}^{\mathrm{expt}}}}        \newc{\amusm}{\ensuremath{ a_{\mu}^{\mathrm{SM}}}}
\newc\deltaamu{\ensuremath{\Delta a_{\mu}}} \newc{\deltaamususy}{\ensuremath{\delta a_{\mu}^{\mathrm{SUSY}}}}
\newc\gmtwo{\ensuremath{ (g-2)_{\mu}}} 
\newc{\deltagmtwomususy}{\ensuremath{\delta\left(g-2\right)_{\mu}^{\mathrm{SUSY}}}}
\newc{\deltagmtwomu}{\ensuremath{\delta\left(g-2\right)_{\mu}}}
\newc\BR{\ensuremath{\rm BR}}
\newc\bsgamma{\ensuremath{ b\rightarrow s \gamma }}
\newc\bxsgamma{\ensuremath{\overline{B}\rightarrow X_{s}\gamma}}
\newc\brbsgamma{\ensuremath{\BR\left(\bsgamma\right)}}
\newc\brbxsgamma{\ensuremath{\BR\left(\bxsgamma\right)}}
\newc\bsmumu{\ensuremath{B_s\to\mu^+\mu^-}}
\newc\brbsmumu{\ensuremath{\BR\left(B_s\to\mu^+\mu^-\right)}}
\newc\bdmmumu{\ensuremath{\overline{B}_d\to\mu^+\mu^-}}
\newc\bbbarmix{\ensuremath{\overline{B}_s\mbox{-}B_s}}      
\newc\delmbs{\ensuremath{\Delta M_{B_s}}}
\newc{\butaunu}{\ensuremath{B_u \rightarrow \tau \nu}}
\newc{\brbutaunu}{\ensuremath{\BR\left(B_u \rightarrow \tau \nu\right)}}

\newcommand*{\reftable}[1]{Table~\ref{#1}}         \newcommand*{\reftables}[2]{Tables~\ref{#1} and \ref{#2}}
       
\newcommand*{\reffig}[1]{Fig.~\ref{#1}}
\newcommand*{\reffigs}[2]{Figs.~\ref{#1} and \ref{#2}} 
\newcommand*{\reffigsd}[2]{Figs.~\ref{#1}-\ref{#2}} 
        \newcommand*{\refeq}[1]{Eq.~\ref{#1}}   

     \newcommand*{\refsec}[1]{Sec.~\ref{#1}}
            
\newcommand*{\refap}[1]{Appendix~\ref{#1}}     
\newcommand*{\refeqst}[2]{Eqs.~\ref{#1}-\ref{#2}}

\newcommand*{\neutone}{\ensuremath{\tilde{\chi}^0_1}}

\newcommand*{\eight}{\ensuremath{\sqrt{s}=8\tev}}

\newcommand*{\madgr}{\texttt{MadGraph5\_aMC@NLO}}

\let\oldcite\cite
\renewcommand*{\cite}{~\oldcite}

\newcommand*{\hl}{\ensuremath{h}}

\newcolumntype{C}[1]{>{\centering\arraybackslash}p{#1}}

\allowdisplaybreaks

\restylefloat{figure}

\title{\bf Road map through the desert with scalars}

\author[a]{Ubaldo Cavazos Olivas,}
\author[b]{Kamila Kowalska}
\author[c]{and Dinesh Kumar}

\affiliation{$^a$ Faculty of Physics, University of Warsaw, Pasteura 5, 02-093 Warsaw, Poland\\
$^b$ National Centre for Nuclear Research,  Pasteura 7, 02-093 Warsaw, Poland\\
$^c$ Department of Physics, University of Rajasthan, Jaipur-302004, India}

\emailAdd{ubaldo.olivas@fuw.edu.pl}
\emailAdd{kamila.kowalska@ncbj.gov.pl}
\emailAdd{dinesh.kumar@ncbj.gov.pl}

\abstract{In the context of the gauge coupling unification, we present a comprehensive analysis of the extensions of the Standard Model with vector-like fermions and scalars. We find 145 models that satisfy the unification condition, which are distinguishable by the number of new particles in the spectrum and by their transformation properties under the gauge symmetry group of the Standard Model. For all models we derive lower bounds on the exotic fermion and scalar masses, stemming from the measurement of the strong gauge coupling scale dependence, from the heavy stable charged particle searches, and from the electroweak precision tests. We also discuss the potential of testing the unification scenarios at the future 100\tev\ collider and in the proton decay experiments. We show that many models can already be excluded based on the current data, while many others will be entirely probed in the coming years.}

\begin{document}
\maketitle
\section{Introduction}\label{sec:intro}

The idea of unification, i.e. the common high-scale origin of three fundamental interactions of the Standard Model (SM), has been driving intellectual efforts in particle physics since the mid 1970s\cite{Georgi:1974sy,Pati:1974yy,Mohapatra:1974hk,Fritzsch:1974nn,Georgi:1974my}. 
The well known observation that the strength of the renormalized gauge couplings of the SM, which significantly differ at the scale of the electroweak (EW) interactions, seems to become similar at very high energies, may suggest the existence of a new unified description of fundamental forces known as a Grand Unified Theory (GUT).

It has also been known for decades that in the standalone SM the precise unification of the gauge couplings is not really achieved, as their GUT-scale values deviate from one another by several percent. This fact provided a strong motivation to consider various extensions of the SM in which the unification is guaranteed by the appropriate choice of the particle spectrum\cite{Rizzo:1991tc,Zhang:2000zy,Choudhury:2001hs,Li:2003zh,Morrissey:2003sc,Giudice:2004tc,ArkaniHamed:2004fb,Dorsner:2005fq,EmmanuelCosta:2005nh,Barger:2006fm,Barger:2007qb,Shrock:2008sb,Calibbi:2009cp,Gogoladze:2010in,Donkin:2010ta,Dermisek:2012as,Dermisek:2012ke,Liu:2012qua,Dorsner:2014wva,Xiao:2014kba,Bhattacherjee:2017cxh,Zheng:2017aaa,Schwichtenberg:2018cka,Zheng:2019kqu,Kowalska:2019qxm,Parida:2020yib}.

In\cite{Kowalska:2019qxm} we proposed to use the concept of gauge coupling unification as an underlying organizing principle for the beyond-the-SM (BSM) model building and to systematically classify the BSM extensions that feature this property. In this way we intended to create a sort of road map that would guide the model building through the desert between the electroweak and the GUT scales. For starters, we looked for all possible extensions of the SM with two types of vector-like (VL) fermions in different representations of the SM gauge group. We found that the precise gauge coupling unification can be achieved for 13 distinct models only, as long as the masses of VL fermions are limited to $10\tev$. We also discussed complementarity of various experimental search strategies that can be employed to probe the BSM spectra favored by the unification requirement. We showed that all but one VL model can be excluded or tested by the current and future experiments. 

The present study is a direct follow up of\cite{Kowalska:2019qxm}. We extend the previous analysis by allowing a BSM theory to contain, beside VL fermions, also extra scalar fields and consider two broad categories of the SM extensions: with one type of fermions plus one type of scalars, and with two types of scalars in different representations. As we will see later, that increases significantly the number of possible unification scenarios as the impact of scalars on the renormalization group equations (RGEs) of the gauge couplings is milder than the one of fermions.

We also perform a detailed phenomenological analysis of all successful unification scenarios. We take into account a wide set of experimental collider searches, which include determination of the strong gauge coupling at different energy scales, measurements of the electroweak precision observables (EWPO), searches for heavy stable charged particles (HSCP), as well as the measurement of the proton lifetime at Super- and Hyper-Kamiokande. We update the analysis of\cite{Kowalska:2019qxm} by incorporating the most recent experimental data and by adding experimental information that were not previously examined. The latter include projections for determination of the running strong and electroweak gauge couplings at a hypothetical future 100\tev\ proton collider, which proves very efficient in testing the parameter space consistent with the unification condition.

The paper is structured as follows. In \refsec{sec:mod} we briefly revise our computational strategy of identifying BSM scenarios that allow for a precise unification of the SM gauge coupling constants. We also present a comprehensive list of all possible models that satisfy this condition. \refsec{sec:tests} is devoted to a detailed discussion of experimental searches that could test, and in many cases exclude, the parameter space of the models. We summarize our results and conclude in \refsec{sec:con}. Technical details of the analysis are collected in four appendices.

\section{BSM models with gauge coupling unification}\label{sec:mod}

In constructing a set of generic anomaly-free extensions of the SM that allow for unification of the three gauge couplings of the SM, we strictly follow the approach introduced in\cite{Kowalska:2019qxm}. First, we define the fundamental building blocks of our models, which are: 

$\bullet$ VL fermions $F$, which transform under the $SU(3)_C\times SU(2)_L\times U(1)_Y$ gauge group as
\be
\mathcal{R}_F=(R_{3F},R_{2F},Y_{F})\oplus(\bar{R}_{3F},\bar{R}_{2F},-Y_{F}),
\ee

$\bullet$ complex scalars $S$ transforming as 
\be
\mathcal{R}_S=(R_{3S},R_{2S},Y_{S}).
\ee
We denote by $N_{F}$ and $N_{S}$ number of copies of VL fermions and scalars that have the same quantum numbers. The left- and right-handed components of a VL pair are counted separately, so $N_{F}$ can only assume even values. The only exception is a fermion transforming in an adjoint representation of the non-abelian gauge symmetry group, in which case $N_{F}$ can be odd. 

In\cite{Kowalska:2019qxm} we discussed in details the SM extensions with two different VL fermion representations. In the present work we extend the previous analysis by incorporating scalar fields. We thus consider fermion plus scalar extensions of the SM, dubbed as `FS' hereafter, which can be entirely parametrized by a set of properties 
\be
\textrm{FS}:\;\mathcal{R}_F,\,\mathcal{R}_S,\,N_{F},\,N_{S},
\ee
and scalar plus scalar extensions, dubbed as `SS', described by
\be
\textrm{SS}:\;\mathcal{R}_{S_1},\,\mathcal{R}_{S_2},\,N_{S_1},\,N_{S_2}.
\ee
At this initial stage both $N_{F}$ and $N_{S_i}$ are unconstrained and it is one of the goals of our analysis to determine their allowed values.

The numerical procedure employed to establish which of the SM extensions  with VL fermions and scalars could potentially lead to the precise unification of three SM gauge couplings at high energies was described in details in\cite{Kowalska:2019qxm} and we briefly summarize it here. For a given BSM model, we initiate the renormalization group (RG) running of the SM gauge couplings at the scale associated with the top quark mass, $M_t$, at which they assume the following values\cite{Buttazzo:2013uya}:
\be\label{init}
g_3(M_t)=1.16660\,,\qquad g_2(M_t)=0.64779\,,\qquad g_Y(M_t)=0.35830.
\ee
We then use the 2-loop SM RGEs given in\cite{Machacek:1983tz} to run the system up from $M_t$ to the scale $M_{1}$, at which the lightest of the BSM particles show up in the spectrum. To simplify the analysis we assume that all VL fermions (complex scalars) in the same representation $\mathcal{R}_{F(S)}$ have a common mass. 
Above the scale $M_1$ we replace the SM RGEs with those corresponding to a generic BSM scenario\cite{Machacek:1983tz}, whose explicit forms adapted to the models considered in this study are given in \refeqst{beta1}{beta3} of Appendix \ref{app:rges}. Similarly, at the scale $M_2$ the impact of heavier BSM states is incorporated and the full system runs up.\footnote{
 For the purpose of this study we additionally assume that all Yukawa interactions generated by the BSM sector and allowed by the gauge symmetry can be neglected.} The unification scale, \mgut, is then defined as the scale at which all three gauge couplings of the SM reach the same value, $g_{\gut}$:
\be
g_{\gut}=g_3(\mgut)=g_2(\mgut)=g_1(\mgut)\,.
\ee
Note that we use the $SU(5)$ normalization, $g_1=\sqrt{5/3}\,g_Y$.

To quantify the precision of the gauge coupling unification we introduce a set of three mismatch parameters, $\epsilon_{1,2,3}$, defined as a deviation of one of the couplings, $g_k$, from the unified value of the remaining two, $g_{ij}=g_i(\mgut^{ij})=g_j(\mgut^{ij})$,
\be
\epsilon_{k}=\frac{g_k^2(\mgut^{ij})-g^2_{ij}}{g^2_{ij}},\quad i,j\neq k.
\ee
The putative unification scale is then calculated by imposing the requirement
\be\label{eq:eps}
\epsilon_{\textrm{GUT}}=\min(\epsilon_1,\epsilon_2,\epsilon_3).
\ee
The last step is to set the upper bound on $\epsilon_{\textrm{GUT}}$, below which the gauge coupling unification will be defined as precise. Of course, any such choice is to certain extent arbitrary. For example, in the SM  $\epsilon_{\textrm{GUT}}^{\textrm{SM}}=7.3\%$, confirming the well known fact that the three gauge couplings do not really unify in this framework. In the minimal SUSY version of the SM, however, $\epsilon_{\textrm{GUT}}$ can be much lower, reaching $\epsilon_{\textrm{GUT}}^{\textrm{MSSM}} = 1.1\%$ if all sparticle masses are set at 1 TeV. We thus believe it is reasonable to define the {\it precise gauge unification} (PGU) by a condition 
\be\label{eq:pgu}
\epsilon_{\textrm{GUT}} \lesssim1\%. 
\ee

Besides the PGU condition, that must be satisfied by any BSM model featuring the gauge coupling unification, there are also other requirements that need to be fulfilled by a model to make it a valid unification scenario. These extra requirements stem both from theoretical consistency of the theory and from the experimental data. We will now discuss them one by one and show that their inclusion significantly reduces the number of viable PGU models.

\paragraph{Perturbativity}

We will require that the running gauge couplings remain perturbative ($g_i\lesssim 4\pi$) up to the energy scale of $10^{15}\gev$, which is the proxy for the proton stability bound to be discussed in more details later on. This condition provides an upper bound on the possible dimensions of representations under which new fermions and scalar transform, and on the numbers of their copies $N_F$ and $N_{S}$. In order to save computation time in the later stages of the analysis, it is convenient to discard all not perturbative models at the very beginning.

\begin{table}[t]\footnotesize
\parbox{.45\linewidth}{
\centering
\begin{tabular}{|c|cccc|}
\hline
\diagbox{$\bm{R_{2F}}$}{$\bm{R_{3F}}$} & $\bm{1}$ & $\bm{3}$ & $\bm{6}$ & $\bm{8}$ \\
\hline
$\bm{1}$ & $\infty$ & 24 & 4 & 4 \\
$\bm{2}$ & 28 & 8 & 2 & 2 \\
$\bm{3}$ & 6 & 2 & 0 & 0 \\
$\bm{4}$ & 2 & 0 & 0 & 0\\
\hline
\end{tabular}
\caption{Maximal number of VL fremions with the common mass of $10\tev$ for which the SM gauge couplings $g_3$ and $g_2$ remain perturbative up to  $10^{15}\gev$. $Y_F=0$ is assumed.}\label{max_num_fer}
}
\hfill
\parbox{.45\linewidth}{
\centering
\begin{tabular}{|c|cccc|}
\hline
\diagbox{$\bm{R_{2F}}$}{$\bm{R_{3F}}$} & $\bm{1}$ & $\bm{3}$ & $\bm{6}$ & $\bm{8}$ \\
\hline
$\bm{1}$ & $3\frac{1}{6}$ & $1\frac{5}{6}$ & $1\frac{1}{6}$ & $1\frac{1}{6}$ \\
$\bm{2}$ & $2\frac{1}{3}$ & $1\frac{1}{3}$ & $\frac{5}{6}$ & $\frac{2}{3}$ \\
$\bm{3}$ & $1\frac{5}{6}$ & 1 & 0 & 0 \\
$\bm{4}$ & $1\frac{2}{3}$ & 0 & 0 & 0 \\
\hline
\end{tabular}
\caption{Maximal value of the hypercharge for which the SM gauge coupling $g_1$ remains perturbative up to  $10^{15}\gev$. ($N_F=2$ and the VL mass is set at $10\tev$).}\label{max_val_hyp}
}
\end{table}

\begin{table}[t]\footnotesize
\parbox{.45\linewidth}{
\centering
\begin{tabular}{|c|cccccc|}
\hline
\diagbox{$\bm{R_{S2}}$}{$\bm{R_{S3}}$} & $\bm{1}$ & $\bm{3}$ & $\bm{6}$ & $\bm{8}$ & $\bm{10}$ & $\bm{15^\prime}$\\
\hline
$\bm{1}$ & $\infty$ & 49 & 8 & 7 & 2 & 1 \\
$\bm{2}$ & 59 & 18 & 4 & 3 & 1 & 0 \\
$\bm{3}$ & 13 & 4 & 1 & 1 & 0 & 0 \\
$\bm{4}$ & 4 & 1 & 0 & 0 & 0 & 0\\
$\bm{5}$ & 2 & 0 & 0 & 0 & 0 & 0\\
$\bm{6}$ & 1 & 0 & 0 & 0 & 0 & 0\\
\hline
\end{tabular}
\caption{Maximal number of complex scalars with the common mass of $10\tev$ for which the SM gauge couplings $g_3$ and $g_2$ remain perturbative up to  $10^{15}\gev$. $Y_S=0$ is assumed.}\label{max_num_sca}
}
\hfill
\parbox{.45\linewidth}{
\centering
\begin{tabular}{|c|cccccc|}
\hline
\diagbox{$\bm{R_{S2}}$}{$\bm{R_{S3}}$} & $\bm{1}$ & $\bm{3}$ & $\bm{6}$ & $\bm{8}$  & $\bm{10}$ & $\bm{15^\prime}$\\
\hline
$\bm{1}$ & $5\frac{1}{6}$ & $3\frac{1}{3}$ & $2\frac{1}{3}$ & 2 & $1\frac{5}{6}$ &  $1\frac{1}{2}$ \\
$\bm{2}$ & $3\frac{5}{6}$ & $2\frac{1}{3}$ & $1\frac{2}{3}$ & $1\frac{1}{2}$ & $1\frac{1}{6}$ & 0 \\
$\bm{3}$ & $3\frac{1}{3}$ & 2 & $1\frac{1}{3}$ & $1\frac{1}{6}$ & 0 & 0 \\
$\bm{4}$ & $2\frac{5}{6}$ & $1\frac{2}{3}$ & 0 & 0 & 0 & 0 \\
$\bm{5}$ & $2\frac{1}{2}$ & 0 & 0 & 0 & 0 & 0\\
$\bm{6}$ & $2\frac{1}{6}$ & 0 & 0 & 0  & 0 & 0 \\
\hline
\end{tabular}
\caption{Maximal value of the hypercharge for which the SM gauge coupling $g_1$ remains perturbative up to  $10^{15}\gev$. ($N_S=1$ and the scalar mass is set at $10\tev$).}\label{max_sca_hyp}
}
\end{table}

In the case of VL fermions, the corresponding perturbativity bounds were derived in\cite{Kowalska:2019qxm} and we recall them here for completeness. In \reftable{max_num_fer} we show the maximal allowed number of VL fermions, $N_{F_{\textrm{max}}}$, for which the SM gauge couplings $g_2$ and $g_3$ remain perturbative up to the unification scale. We set $Y_F=0$ (non-zero hypercharge could only reduce $N_{F_{\textrm{max}}}$) and perform the analysis for different combinations of representations $R_{3F}$ and $R_{2F}$. To be conservative and to avoid excluding viable scenarios, we fix the VL mass at $10\tev$, which is the largest allowed mass scale we consider in this study.  In \reftable{max_val_hyp} we report instead the maximal value of the hypercharge allowed by perturbativity of $g_1$ for a single pair of VL fermions ($N_{F}=2$). 

The corresponding results for complex scalars are presented in \reftable{max_num_sca} and \reftable{max_sca_hyp}. Larger representations are allowed in this case due to the fact that scalar contributions to the gauge beta functions are by a factor of two smaller than the fermion ones, c.f. \refeqst{oneloop3}{oneloop1}. We will see later that this property results in a larger number of the allowed PGU scenarios with at least one BSM scalar in comparison to the two-fermion case.

\paragraph{$\bm{SU(5)}$ unification} Following\cite{Kowalska:2019qxm}, we assume that all BSM fields are embedded at the unification scale into the $SU(5)$ multiplets. The decomposition of the irreducible $SU(5)$ representations into irreducible representations of the SM gauge group is presented in \refap{su5}. Note that the representations of the dimension larger than 75 can be decomposed either to representations that have already appeared in the decomposition of lower dimensional representations, or to representations whose dimensions exceed the limits presented in \reftables{max_num_fer}{max_num_sca}. Taking into account hypercharge perturbativity bounds from \reftables{max_val_hyp}{max_sca_hyp}, we are left with 24 fermion and 34 scalar distinct non-singlet $SU(3)_C\times SU(2)_L\times U(1)_Y$ representations:

\begin{itemize}
\item color singlets: $\left({\bf 1}, {\bf 1}, 1\right)$, $\left({\bf 1}, {\bf 1}, -2\right)$, $\left({\bf 1}, {\bf 2}, \textstyle\frac{1}{2}\right)$, $\left({\bf 1}, {\bf 2}, -\textstyle\frac{3}{2}\right)$, $\left({\bf 1}, {\bf 3}, 0\right)$, $\left({\bf 1}, {\bf 3}, 1\right)$, $\left({\bf 1}, {\bf 4}, \textstyle\frac{1}{2}\right)$, $\left({\bf 1}, {\bf 4}, -\textstyle\frac{3}{2}\right)$, $\left({\bf 1}, {\bf 5}, -2\right)^S$,

\item color triplets: $\left({\bf 3}, {\bf 1}, -\textstyle\frac{1}{3}\right)$, $\left({\bf\bar{3}}, {\bf 1}, -\textstyle\frac{2}{3}\right)$, $\left({\bf\bar 3}, {\bf 1}, \textstyle\frac{4}{3}\right)$, $\left({\bf\bar 3}, {\bf 1}, -\textstyle\frac{5}{3}\right)$,  $\left({\bf 3}, {\bf 2}, \textstyle\frac{1}{6}\right)$, $\left({\bf\bar 3} , {\bf 2}, \textstyle\frac{5}{6}\right)$, $\left({\bf\bar 3}, {\bf 2}, -\textstyle\frac{7}{6}\right)$, $\left({\bf 3}, {\bf 3}, -\textstyle\frac{1}{3}\right)$, $\left({\bf\bar 3}, {\bf 3}, -\textstyle\frac{2}{3}\right)$, $\left({\bf\bar 3}, {\bf 3}, \textstyle\frac{4}{3}\right)^S$, $\left({\bf\bar 3}, {\bf 4}, -\textstyle\frac{7}{6}\right)^S$,

\item color sextets: $\left({\bf\bar 6}, {\bf 1}, -\textstyle\frac{1}{3}\right)$, $\left({\bf 6}, {\bf 1}, -\textstyle\frac{2}{3}\right)$, $\left({\bf 6}, {\bf 1}, \textstyle\frac{4}{3}\right)^S$, $\left({\bf\bar 6}, {\bf 2}, \textstyle\frac{1}{6}\right)$, $\left({\bf 6}, {\bf 2}, \textstyle\frac{5}{6}\right)$, $\left({\bf 6}, {\bf 2}, -\textstyle\frac{7}{6}\right)^S$, $\left({\bf\bar 6}, {\bf 3}, -\textstyle\frac{1}{3}\right)^S$,

\item color octets: $\left(\bm{8}, \bm{1}, 0\right)$, $\left(\bm{8}, \bm{1}, 1\right)$, $\left({\bf 8}, {\bf 2}, \textstyle\frac{1}{2}\right)$, $\left(\bm{8}, \bm{3}, 0\right)^S$,
\item higher dimensions: $\left({\bf\overline{10}}, {\bf 1}, 1\right)^S$, $\left({\bf \overline{10}}, {\bf 2}, \textstyle\frac{1}{2}\right)^S$, $\left({\bf\overline{15}^{\prime}}, {\bf 1}, \textstyle\frac{4}{3}\right)^S$,
\end{itemize}
where the superscript $S$ indicates that a given representation is allowed for the scalars only. In total there are 816 distinct perturbative models of type FS and 561 distinct perturbative models of type SS for which the possibility of precise gauge coupling unification needs to be investigated.

\paragraph{Proton decay}

Proton decay is a generic prediction of the GUT theories. In the framework of $SU(5)$ the SM quarks and leptons belong to the same $SU(5)$ multiplets. As a result, new heavy gauge bosons can mediate the interactions that violate both the baryon and lepton number conservation.  The dominant contribution to the proton decay width stems from the dimension-6 operators of the type $QQQL$. While the exact form of these countributions is highly model-dependent, the approximate proton lifetime can be extracted from the unification scale \mgut\ and the value of the unified gauge coupling $g_{\textrm{GUT}}$ as\cite{Nath:2006ut}
\be\label{life}
\tau_p=\left(\frac{4\pi}{g^2_{\textrm{GUT}}}\right)^2\left(\frac{\mgut}{\gev}\right)^4\times 2.0\times 10^{-32}\;\textrm{years}.
\ee

Proton decay has been experimentally tested since the early 1990s by the underground water Cherenkov detector Super-Kamiokande (SK). The strongest lower bound on the proton lifetime comes from the decay channel $p\to e^+\pi^0$ and is given by\cite{Miura:2016krn} 
\be\label{eq.prot}
\tau_p>1.6\times 10^{34}\;\; \textrm{years}.
\ee 
It is expected that in the not-so-far future the limit will be extended by at least one order of magnitude, up to $\sim 2\times 10^{35}$ years\cite{Abe:2018uyc}, by a new generation detector Hyper-Kamiokande (HK).

We apply the SK bound of \refeq{eq.prot} as an extra viability condition and discard all the models that do not satisfy it.

\subsection{List of the allowed models}

For each of 816 distinct models of type FS and 561 distinct models of type SS  we scan over the number of BSM fermions and/or scalars, $N_F,\,N_S$, and their common masses, $M_1,\,M_2$. The upper bounds on the number of BSM particles are given in \reftables{max_num_fer}{max_num_sca}, while the parameters $M_1$ and $M_2$ range from 0.25\tev\ to 10\tev. For each point in the 4-dimensional parameter space we determine $\epsilon_{\textrm{GUT}}$ and $\mgut$ from \refeq{eq:eps} and verify whether the PGU condition in \refeq{eq:pgu} is satisfied. Finally, we check if the experimental bound on the proton lifetime, \refeq{eq.prot}, is not exceeded.

The results of our numerical analysis are summarized in \reftable{2reps_all_comb}. In total, we found 22 different FS scenarios that allow for the PGU at the scale $10^{15}-10^{18}\gev$. The corresponding quantum numbers of fermions and scalar w.r.t. the $SU(3)_C\times SU(2)_L\times U(1)_Y$ gauge group  are shown in columns 2 and 3. Many of the successful scenarios allow for various numbers of fermion and scalar copies, which are indicated in the fourth column of \reftable{2reps_all_comb}. If these two extra degrees of freedom are taken into account, one can identify 67 distinct FS-type PGU models, which we will denote as FSI$^{(N_F,N_S)}$, with I$=1,\dots, 22$.

\begin{scriptsize}
\begin{center}
\begin{longtable}[t]{|c|cc|c|}
\hline
Scenario & $\bm{R_{F}}$ & $\bm{R_{S}}$ & $(N_F,\,N_S)$  \\
\hline
FS1 & $\left(\bm{1}, \bm{2}, \frac{1}{2}\right)$ & $\left({\bf 6}, \bm{1}, \frac{1}{3}\right)$ & $(16/18/20,\,6)$, $(22,\,7)$ \\
FS2 & $\left(\bm{1}, \bm{2}, \frac{1}{2}\right)$ & $\left({\bf 8}, \bm{2}, \frac{1}{2}\right)$ & $(4,\,2)$ \\
FS3 & $\left(\bm{1}, \bm{3}, 0\right)$ & $\left({\bf 3}, \bm{1}, -\frac{1}{3}\right)$ & $(2,\,13/14/15/16/17)$,  $(3,\,22/23/24/25/26/27)$  \\
FS4 & $\left(\bm{1}, \bm{3}, 0\right)$ & $\left({\bf 6}, \bm{1}, \frac{1}{3}\right)$ & $(2,\,3)$\\
FS5 & $\left(\bm{1}, \bm{3}, 0\right)$ & $\left({\bf 6}, \bm{1}, -\frac{2}{3}\right)$ & $(3,\,4)$, $(4,\,5/6)$, $(5,\,7)$ \\
FS6 & $\left(\bm{3}, \bm{1}, -\frac{1}{3}\right)$ & $\left({\bf 1}, \bm{4}, \frac{1}{2}\right)$ & $(10/12,\,2)$  \\
FS7 & $\left(\bm{3}, \bm{1}, -\frac{1}{3}\right)$ & $\left({\bf 3}, \bm{2}, \frac{1}{6}\right)$ & $(2,\,4)$, $(4,\,6)$ \\
FS8 & $\left(\bm{3}, \bm{1}, -\frac{1}{3}\right)$ & $\left({\bf 3}, \bm{3}, -\frac{1}{3}\right)$ & $(4,\,1)$ \\
FS9 & $\left(\bm{3}, \bm{1}, -\frac{1}{3}\right)$ & $\left({\bf 3}, \bm{3}, \frac{2}{3}\right)$ &  $(10,\,2)$, $(16,\,3)$\\
FS10 & $\left(\bm{3}, \bm{1}, -\frac{2}{3}\right)$ & $\left({\bf 3}, \bm{2}, \frac{1}{6}\right)$ & $(2,\,5)$, $(4,\,7/8)$, $(6,\,10/11)$ \\
FS11 & $\left(\bm{3}, \bm{1}, -\frac{2}{3}\right)$ & $\left({\bf 6}, \bm{3}, \frac{1}{3}\right)$ & $(6,\,1)$ \\
FS12 & $\left(\bm{3}, \bm{2}, \frac{1}{6}\right)$ & $\left({\bf 3}, \bm{1}, -\frac{1}{3}\right)$ & $(2,\,2/3/4)$ \\
FS13 & $\left(\bm{3}, \bm{2}, \frac{1}{6}\right)$ & $\left({\bf 3}, \bm{1}, -\frac{2}{3}\right)$ & $(4,\,7/8)$, $(6,\,11/12/13)$ \\
FS14 & $\left(\bm{3}, \bm{2}, \frac{1}{6}\right)$ & $\left({\bf 6}, \bm{2}, \frac{5}{6}\right)$ & $(4,\,2)$ \\
FS15 & $\left(\bm{3}, \bm{2}, \frac{1}{6}\right)$ & $\left({\bf 8}, \bm{1}, 1\right)$ & $(4,1)$, $(6,\,2)$ \\
FS16 & $\left(\bm{6}, \bm{1}, \frac{1}{3}\right)$ & $\left({\bf 1}, \bm{2}, \frac{1}{2}\right)$ & $(4,\,39/40/41/42/43/44/45/46/47)$\\
FS17 & $\left(\bm{6}, \bm{1}, \frac{1}{3}\right)$ & $\left({\bf 3}, \bm{3}, \frac{2}{3}\right)$ & $(2,\,2)$ \\
FS18 & $\left(\bm{6}, \bm{1}, -\frac{2}{3}\right)$ & $\left({\bf 1}, \bm{4}, \frac{1}{2}\right)$ & $(4,\,4)$ \\
FS19 & $\left(\bm{6}, \bm{1}, -\frac{2}{3}\right)$ & $\left({\bf 3}, \bm{3}, -\frac{1}{3}\right)$ & $(2,\,2)$ \\
FS20 & $\left(\bm{8}, \bm{1}, 0\right)$ & $\left({\bf 1}, \bm{2}, \frac{1}{2}\right)$ & $(2,\,23/24/25/26/27/28)$, $(3,\,35/36)$  \\
FS21 & $\left(\bm{8}, \bm{1}, 0\right)$ & $\left({\bf 1}, \bm{4}, \frac{1}{2}\right)$ & $(1,\,1)$ \\
FS22 & $\left(\bm{8}, \bm{1}, 0\right)$ & $\left({\bf 3}, \bm{3}, -\frac{1}{3}\right)$ & $(1,\,1)$ \\
\hline
Scenario & $\bm{R_{S_1}}$ & $\bm{R_{S_2}}$ & $(N_{S_1},\,N_{S_2})$  \\
\hline
SS1 & $\left(\bm{1}, \bm{2}, \frac{1}{2}\right)$ & $\left({\bf 6}, \bm{1}, \frac{1}{3}\right)$ & $(40/41/42,\,6)$, $(43/44/45/46/47,\, 7)$ \\
SS2 & $\left(\bm{1}, \bm{2}, \frac{1}{2}\right)$ & $\left({\bf 8}, \bm{2}, \frac{1}{2}\right)$ & $(8/9/10,\,2)$ \\
SS3 & $\left(\bm{1}, \bm{4}, \frac{1}{2}\right)$ & $\left({\bf 3}, \bm{1}, -\frac{1}{3}\right)$ & $(2,\,18/19/20/21/22/23/24)$ \\
SS4 & $\left(\bm{1}, \bm{4}, \frac{1}{2}\right)$ & $\left({\bf 6}, \bm{1}, -\frac{2}{3}\right)$ & $(4,\, 7)$ \\
SS5 & $\left(\bm{3}, \bm{1}, -\frac{1}{3}\right)$ & $\left({\bf 3}, \bm{2}, \frac{1}{6}\right)$ & $(3/4/5,\, 4)$, $(5/6/7,\, 5)$,  $(7/8,\, 6)$ \\
SS6 & $\left(\bm{3}, \bm{1}, -\frac{1}{3}\right)$ & $\left({\bf 3}, \bm{3}, \frac{2}{3}\right)$ & $(27/28/29/30/31/32/33/34,\,3)$ \\
SS7 & $\left(\bm{3}, \bm{1}, -\frac{2}{3}\right)$ & $\left({\bf 3}, \bm{2}, \frac{1}{6}\right)$ & $(5/6,\, 6)$, $(6/7/8,\, 7)$, $(7/8/9,\, 8)$, $(8/9/10/11,\, 9)$, $(10/11/12,\, 10)$,  \\
& & & $(11/12/13/14,\,11)$, $(13/14/15,\, 12)$, $(14/15,\, 13)$ \\
SS8 & $\left(\bm{3}, \bm{1}, -\frac{2}{3}\right)$ & $\left({\bf 6}, \bm{3}, \frac{1}{3}\right)$ & $(9/10/11/12,\,1)$ \\
SS9 & $\left(\bm{3}, \bm{2}, \frac{1}{6}\right)$ & $\left({\bf 6}, \bm{1}, \frac{1}{3}\right)$ & $(4/5,\, 1)$ \\
SS10 & $\left(\bm{3}, \bm{2}, \frac{1}{6}\right)$ & $\left({\bf 6}, \bm{1}, -\frac{2}{3}\right)$ &  $(5,\, 1)$\\
SS11 & $\left(\bm{3}, \bm{2}, \frac{1}{6}\right)$ & $\left({\bf 6}, \bm{2}, \frac{5}{6}\right)$ & $(5,\, 1)$, $(7/8,\,2)$ \\
SS12 & $\left(\bm{3}, \bm{2}, \frac{1}{6}\right)$ & $\left({\bf 8}, \bm{1}, 1\right)$ &  $(7,\, 1)$,  $(11/12,\,2)$\\
SS13 & $\left(\bm{3}, \bm{2}, \frac{1}{6}\right)$ & $\left({\bf 8}, \bm{2}, \frac{1}{2}\right)$ & $(3,\, 1)$ \\
SS14 & $\left(\bm{3}, \bm{3}, -\frac{1}{3}\right)$ & $\left({\bf 6}, \bm{1}, \frac{1}{3}\right)$ & $(1,\, 2)$ \\
SS15 & $\left(\bm{3}, \bm{3}, -\frac{1}{3}\right)$ & $\left({\bf 6}, \bm{1}, -\frac{2}{3}\right)$ & $(2,\, 3/4)$ \\
SS16 & $\left(\bm{3}, \bm{3}, \frac{2}{3}\right)$ & $\left({\bf 6}, \bm{1}, \frac{1}{3}\right)$ & $(2,\, 4)$ \\
SS17 & $\left(\bm{1}, \bm{3}, 1\right)$ & $\left({\bf 8}, \bm{2}, \frac{1}{2}\right)$ & $(3,\, 3)$ \\
\hline
\caption{FS and SS extensions of the SM, which allow for the  PGU ($\epsilon_{\textrm{GUT}}\leq 1\%$) at the unification scale $10^{15}-10^{18}\gev$ and are not excluded by the SK measurement of the proton lifetime\cite{Miura:2016krn}. The BSM masses lie in the range $0.25-10\tev$. In columns 2 and 3 quantum numbers of the new particles w.r.t.~the SM gauge group are given. Column 4 displays possible numbers of identical copies of fermions and scalars in each representation. 
}
\label{2reps_all_comb}
\end{longtable}
\end{center}
\end{scriptsize}

In the case of SS-type models, 17 distinct PGU scenarios were found. Note that the gauge coupling unification is not possible with the representations of a dimension higher than 8, even though these are allowed by perturbativity condition. Moreover, PGU  can not be achieved with scalars in the adjoint representations of either $SU(3)_C$ or $SU(2)_L$, contrarily to what is observed for VL fermions. Taking into account different numbers of scalar copies corresponding to the same representation, we can define 78 distinct models denoted as SSJ$^{(N_{S_1},N_{S_2})}$, with J$=1,\dots, 17$. They are all listed in \reftable{2reps_all_comb}.\footnote{The full list of two-fermion (FF) extensions of the SM was derived in\cite{Kowalska:2019qxm} and listed in Table 3 therein. We recall that only 7 possible combinations of representations were allowed by the PGU condition, corresponding to 13 distinct models with different numbers of fermion copies.} 

\begin{figure}[t]
\centering
\subfloat[Type H0 - model SS6$^{(31,3)}$]{%
\includegraphics[width=0.33\textwidth]{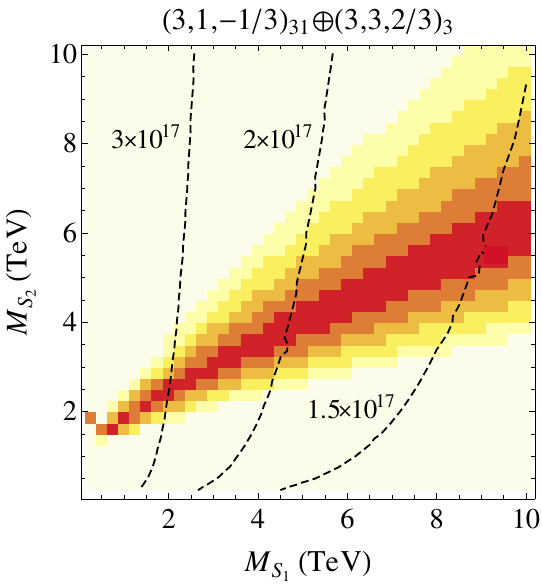}
}%
\hspace{0.1cm}
\subfloat[Type H1 - model SS3$^{(2,19)}$]{%
\includegraphics[width=0.33\textwidth]{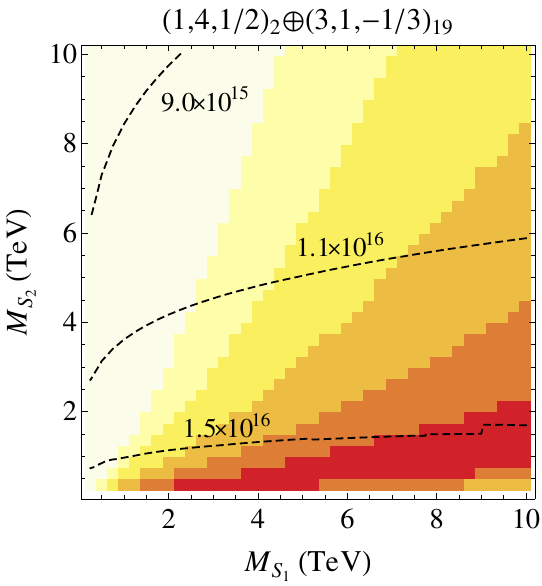}
}%
\hspace{0.0cm}
 \raisebox{0.5\height}{\includegraphics[width=0.03\textwidth]{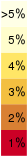}}
\\
\vspace{0.5cm}
\subfloat[Type H2 - model SS7$^{(10,10)}$]{%
\includegraphics[width=0.33\textwidth]{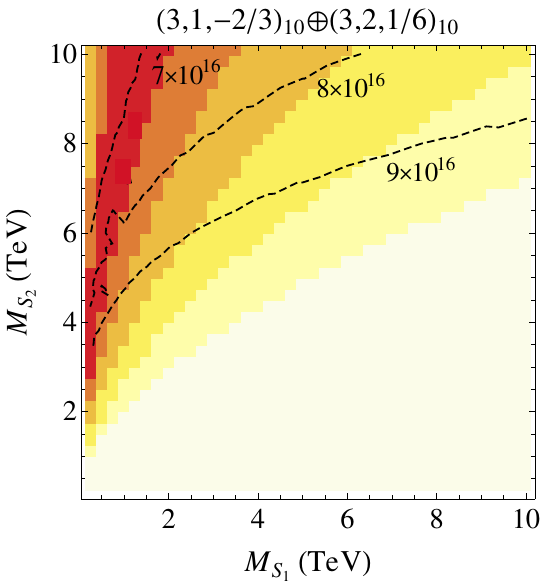}
}%
\hspace{0.1cm}
\subfloat[Type H3 - model SS10$^{(5,1)}$]{%
\includegraphics[width=0.33\textwidth]{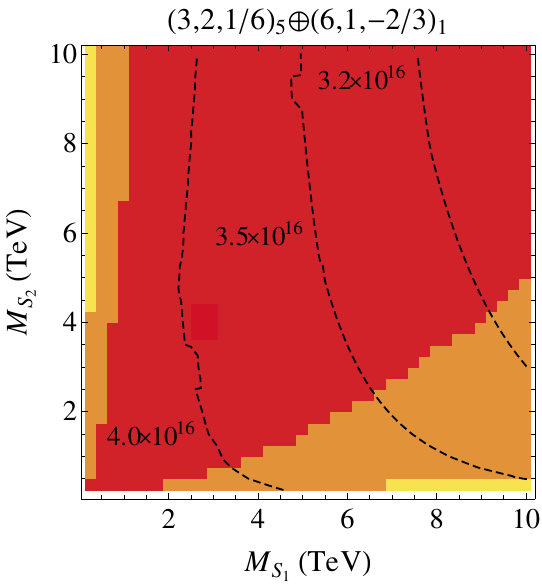}
}%
\hspace{0.0cm}
 \raisebox{0.5\height}{\includegraphics[width=0.03\textwidth]{Figs/legend.png}}
\caption{\footnotesize Examples of various types of the BSM spectra observed in the unification models. Shown is distribution of the mismatch parameter $\epsilon_{\textrm{GUT}}$ as a function of $M_{S_1}$ and $M_{S_2}$ for selected SSJ$^{(N_{S_1},N_{S_2})}$ scenarios listed in \reftable{2reps_all_comb}. PGU region ($\epsilon_{\textrm{GUT}}\leq 1\%$) is marked in red. Different shades of yellow illustrate departure from the precise unification condition as quantified by the increasing values of $\epsilon_{\textrm{GUT}}$. Dashed black contours indicate the unification scale (in \gev). 
\label{fig:models1}}
\end{figure}

The unification requirement does not only select specific representations of the SM gauge symmetry groups under which the BSM particles should transform, but it also determines the spectrum, i.e.~a particular hierarchy between the mass parameters $M_1$ and $M_2$. In this regard, the successful PGU scenarios can be divided in four categories, which will be referred to with the following labels hereafter:
\bea\label{mass_hier}
\rm{H0}:& \;M_{1}\sim M_{2}\,,&\qquad{\rm degenerate\;\; spectrum},\nonumber\\
\rm{H1}:& \;M_{1}\gg M_{2}\,,&\qquad{\rm hierarchical\;\; spectrum},\nonumber\\
\rm{H2}:& \;M_{1}\ll M_{2}\,,&\qquad{\rm hierarchical\;\; spectrum},\nonumber\\
\rm{H3}:& \qquad{\rm unconstrained\;\; spectrum}.
\eea
An illustration of different types of the PGU spectra is presented in \reffig{fig:models1}. Each of the four panels shows a distribution of the mismatch parameter $\epsilon_{\textrm{GUT}}$ as a function of the parameters  $M_1$ and $M_2$ for a specific SSJ$^{(N_{S_1},N_{S_2})}$ scenario from \reftable{2reps_all_comb}. The PGU region, in which $\epsilon_{\textrm{GUT}}<1\%$,  is marked in red. Different shades of yellow indicate departure from the precise unification condition as quantified by the increasing values of $\epsilon_{\textrm{GUT}}$, whose upper bound is given in the legend bar on the right hand side of the plot. Dashed black contours indicate the unification scale \mgut\ (in \gev). Types of spectra characteristic of a given PGU scenario are also indicated in the second column of \reftables{2reps_FS}{2reps_SS} in Appendix \ref{app:list}.

In\cite{Kowalska:2019qxm} we demonstrated that a particular hierarchy among the mass parameters required by the unification condition allows one to probe many of the PGU scenarios of the type FF up to at least 10 TeV thanks to complementarity of various experimental search strategies. In the present study we will assume the same approach to derive experimental bounds on the PGU models listed in \reftable{2reps_all_comb}.   

\section{Testing the PGU models at the colliders}\label{sec:tests}

In general, there are two broad categories of experimental searches that one may want to employ to probe the PGU models identified in \refsec{sec:mod}. The first category encompasses those searches that provide the exclusion bounds which are model independent, in the sense that once the gauge properties of a BSM scenario are defined (in other words, once a particular model from \reftable{2reps_all_comb} is chosen), the experimental limits would not depend on further assumptions regarding the model properties, like the presence of other (non-gauge) interactions. Contrarily, the predictions based on the searches belonging to the second category strongly hinge on those additional  assumptions. In particular, in \refsec{ssec:dep} the experimental lower bounds on fermion and scalar masses will be derived under the assumption that the Yukawa interactions between the BSM and the SM particles are negligible. 

A similar analysis was performed in\cite{Kowalska:2019qxm} for the FF-type PGU models. In this work we extend the previous study to the scenarios of FS and SS type. Additionally, we update some of the experimental searches used in\cite{Kowalska:2019qxm} to incorporate the more recent results. Finally, we discuss the projections derived from  putative measurements of the running weak and strong gauge coupling constants at the future 100\tev\ collider.

\subsection{Model-independent tests}\label{sub:in}
\paragraph{Running of the strong gauge coupling}

The deviation of the renormalized strong gauge coupling constant from its SM running is determined by the representation and the number of  BSM particles that transform non-trivially under the $SU(3)_C$ gauge group. At one-loop the effect is controlled by the size of corrections to the coefficient $B_3$  of the corresponding beta function, \refeq{beta1}, which reads
\be
\Delta B_3\equiv B_3-B_3^{\textrm{SM}}=\frac{2}{3}N_F\,S(R_{F3})\,d(R_{F2})+\frac{1}{3}N_S\,S(R_{S3})\,d(R_{S2}).
\ee 
 
Last year the ATLAS collaboration released a new determination of the running $g_3(Q)$, based on the measurements of the transverse energy-energy correlations (TEEC) and their associated azimuthal asymmetric (ATEEC) functions, using a data sample containing 139 fb$^{-1}$ of proton-proton collisions at the center-of-mass energy of $\sqrt{s} = 13$\,TeV\cite{ATLAS-CONF-2020-025}. The new determination of $g_3(Q)$ shows a very good agreement with the SM predictions up to the energy scale of about $4$\,TeV. These results updated the previous ATLAS analysis of the same type, based on the $\sqrt{s} = 8$\,TeV data set\cite{Aaboud:2017fml}, in which the running of the strong coupling constant was tested up to around $1$\, TeV; and the strong coupling determination based on the measurement of the double-differential inclusive jet cross section at \eight\ by CMS\cite{Khachatryan:2016mlc}, in which the consistency of the $g_3(Q)$ with the SM prediction was confirmed up to $1.5\tev$. 

The experimental searches measuring the energy dependence of $g_3$ use as the main kinematical variable the leading jet transverse momentum $p_T$. As a result, the maximal energy scale at which the running coupling can be evaluated is limited by the value of $p_T$. According to the report\cite{Mangano:2016jyj}, which summarizes most important properties of SM processes at the future 100\tev\ proton collider, the reach in $p_T$ at such a machine could extend well above 20\tev, assuming the total integrated luminosity of 1 ab$^{-1}$. To derive a (very approximate) projection for the future impact of the running strong coupling measurement on our models, we will assume that $g_3$ does not deviate from the SM prediction up to at least 20\tev\ and will use this number as a rough approximation for the maximal energy scale at which it can be probed.

\begin{figure}[t]
\centering
\subfloat[Excluded by $g_3$ - model SS1$^{(40,6)}$]{%
\includegraphics[width=0.37\textwidth]{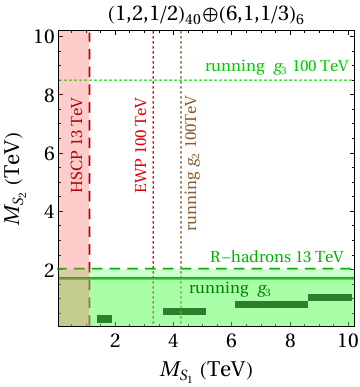}
}%
\hspace{0.3cm}
\subfloat[Testable by $g_3$ - model FS6$^{(10,2)}$]{%
\includegraphics[width=0.37\textwidth]{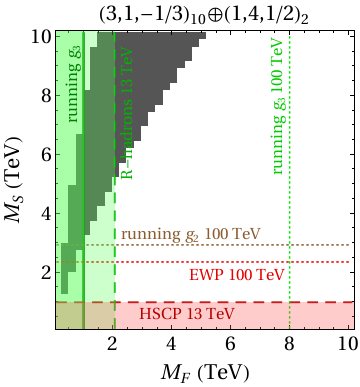}
}%
\caption{\footnotesize Summary of experimental bounds on masses $M_1$ and $M_2$ in scenarios (a) SS1$^{(40,6)}$ and (b) FS6$^{(10,2)}$. In dark gray the PGU region is indicated. The area below and left to the solid green line is excluded by the measurement of the running strong coupling constant by ATLAS\cite{ATLAS-CONF-2020-025}. The limits from the 13 TeV ATLAS $R$-hadrons search\cite{Aaboud:2019trc} are indicated as a dashed green line. The corresponding lepton-like HSCP search excludes the area left and below to a dashed red line. 100 TeV collider projections for the running $g_3$\cite{Mangano:2016jyj}, running $g_2$\cite{Alves:2014cda}, and for the EWP tests\cite{Farina:2016rws} are depicted as green, brown and red dotted lines, respectively. 
\label{fig:bounds1}}
\end{figure}

The impact of the strong coupling determination on the PGU scenarios, derived using the combination of ATLAS measurements with $\sqrt{s} = 8$\,TeV and $\sqrt{s} = 13$\,TeV, is illustrated in \reffig{fig:bounds1} for two exemplary cases that are either (a) excluded by the measurement of $g_3$ (up to $10\tev$), or (b) can be tested by this measurement in the future. The corresponding exclusion bounds are indicated as green solid lines, while the 100\tev\ projections as green dotted lines. Dark gray shaded area corresponds to the PGU region of the parameter space, $\epsilon_{\textrm{GUT}}\lesssim 1\%$. The meaning of the remaining lines will be explained later on.

The lower bounds on the masses of color particles in all the PGU models listed in \reftable{2reps_all_comb} are also collected in column 3 of \reftables{2reps_FS}{2reps_SS} in Appendix \ref{app:list}. Note that in our previous study\cite{Kowalska:2019qxm} the CMS search\cite{Khachatryan:2016mlc} was employed, and thus the present exclusion bounds are significantly stronger. The corresponding lower bounds based on the 100\tev\ projection are shown in column 9 of \reftables{2reps_FS}{2reps_SS}. As it was noted in\cite{Kowalska:2019qxm}, the running of $g_3$ is particularly efficient in testing the hierarchical PGU spectra in which the color particle in much lighter than the non-color one, allowing one to exclude, or to probe in the future, many of the models characterized by the spectrum type H1 or H2. 

\paragraph{Running of the electroweak gauge couplings}

A similar strategy could in principle be applied to the electroweak couplings $g_1$ and $g_2$. While the running of neither of them has been experimentally measured yet, in\cite{Alves:2014cda} it was proposed  to look at the high invariant mass spectrum of the Drell-Yan (DY) production $p\,p\to\gamma/Z^*\to l^+l^-$ measured by the LHC to derive the limits on the energy dependence of $g_1$ and $g_2$. Since the available experimental data from LEP and the 7\tev\ LHC run did not produce strong constraints,\cite{Alves:2014cda} provided predictions for 8, 14 and 100\tev. By now, the actual neutral current DY cross section was measured by CMS\cite{CMS:2014jea} and ATLAS\cite{Aad:2016zzw} at 8\tev, and by CMS at 13\tev\cite{CMS:2018mdl} with 2.3 fb$^{-1}$ of data, finding good agreement of the measured differential cross sections with the theoretical calculations. Therefore, we will treat the 8\tev\ results of\cite{Alves:2014cda} as the actual limits.
 
The DY constraints on $g_1$ and $g_2$ at a given energy scale can be translated into the limits on one-loop BSM contributions to the corresponding beta functions,  
\bea\label{eq:DeltaEW}
\Delta B_2\equiv B_2-B_2^{\textrm{SM}}&=&\frac{2}{3}N_F\,S(R_{F2})\,d(R_{F3})+\frac{1}{3}N_S\,S(R_{S2})\,d(R_{S3}),\\
\Delta B_1\equiv B_1-B_1^{\textrm{SM}}&=&\frac{2}{5} N_F\, d(R_{F3}) d(R_{F2}) Y_F^2+\frac{1}{5} N_S\, d(R_{S3}) d(R_{S2}) Y_S^2,\nonumber
\eea
as a function of the BSM mass. Since the process $p\,p\to\gamma/Z^*\to l^+l^-$ is sensitive to both $g_1$ and $g_2$, the limits in\cite{Alves:2014cda} are provided separately on $\Delta B_2$ under the assumption that there is no BSM contribution to the running of the hypercharge coupling ($\Delta B_1=0$), and on $\Delta B_1$ assuming that $\Delta B_2=0$. We applied those limits to all the PGU models listed in \reftable{2reps_all_comb} and found  that none of them is presently constrained by the DY production.

We can use, however, the results of\cite{Alves:2014cda} to quantify the future reach of the $g_2$ measurement  at the future collider with the centre-of-mass energy of 100\tev\ and 3 ab$^{-1}$ of data. The strongest exclusion bounds can be derived in this case from the charged current DY process $p\,p\to W^*\to l\nu$, which also makes it possible to constrain $\Delta B_2$ individually, with no further assumptions regarding the $U(1)_Y$ charges. The corresponding projected lower bounds on the BSM fermion and scalar masses in our PGU models are reported in column 8 of \reftables{2reps_FS}{2reps_SS}.

\begin{figure}[t]
\centering
\subfloat[Testable by $g_2$ - model FS3$^{(3,27)}$]{%
\includegraphics[width=0.37\textwidth]{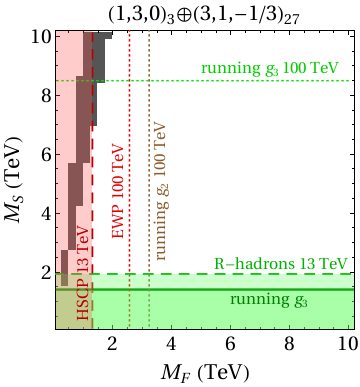}
}%
\hspace{0.3cm}
\subfloat[Testable by $g_2$ - model FS16$^{(4,41)}$]{%
\includegraphics[width=0.37\textwidth]{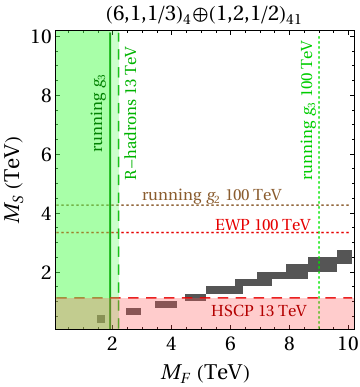}
}%
\caption{\footnotesize Summary of experimental bounds on masses $M_1$ and $M_2$ for scenarios (a) FS3$^{(3,27)}$ and (b) FS16$^{(4,41)}$. The color code is the same as in \reffig{fig:bounds1}.
\label{fig:bounds2}}
\end{figure}

In \reffig{fig:bounds2} we present two PGU scenarios whose parameter space could be entirely tested by the future measurement of $g_2$ in the charged current DY process. The corresponding lower bounds on fermion and scalar masses are indicated as dotted brown lines. One observes that the $g_2$ limits are complementary to the limits form the strong gauge coupling measurement and could provide a constraint on those models in which non-color particles are lighter than the color ones. The bounds, however, are significantly weaker due to the lower production cross section of the EW gauge bosons at hadron colliders. 
 
\paragraph{Electroweak precision tests}

Bounds on the BSM sector can also be derived by looking at the DY processes below the BSM mass threshold and by testing deviations from the SM predictions in the EWPO\cite{Farina:2016rws}. The oblique parameters sensitive to the presence of states charged under the EW gauge symmetry are $W$ and $Y$\cite{Barbieri:2004qk,Cacciapaglia:2006pk}, related to the corresponding beta functions as\cite{Alves:2014cda}
\begin{align}\label{cwyone}
W,Y 
=\frac{g_{2,1}^2}{80\pi^2} \frac{m_W^2}{M_{\textrm{BSM}}^2} \times \Delta B_{2,1} \, ,
\end{align}
where $\Delta B_{2,1}$ are given in \refeq{eq:DeltaEW}.
 
In\cite{Farina:2016rws} the limits on $W$ and $Y$ were derived based on the data from LEP\cite{Falkowski:2015krw}, and from the LHC 8 TeV measurements by ATLAS\cite{Aad:2016zzw} and CMS\cite{CMS:2014jea}. We checked that they do not provide any bounds on the parameter space of the PGU scenarios from \reftable{2reps_all_comb}. However, as it was the case for the running coupling $g_2$, the present experimental bounds can be improved at the projected 100 TeV machine with the luminosity 3 ab$^{-1}$ by roughly two orders of magnitude\cite{Farina:2016rws}. The corresponding projections are summarized in column 7 of \reftables{2reps_FS}{2reps_SS}. They are also indicated as dotted red lines in \reffigs{fig:bounds1}{fig:bounds2}. 

Both the EWPO and running $g_2$ bounds are particularly powerful in constraining the BSM scalars and fermions that transform non-trivially under $SU(2)_L$, and thus they can probe the PGU spectra characterized by the non-color particle lighter than the color one. It is worth to emphasize that all three model-independent searches discussed in this subsection provide a set of complementary constraints that proves to be particularly efficient in testing the hierarchical spectra of the H1 and H2 type. As a matter of fact, a close inspection of  \reftables{2reps_FS}{2reps_SS} reveals that these are exactly the PGU models that are excluded or testable by the model-independent collider searches.

\paragraph{Proton lifetime}

Although technically not being a collider search, the proton decay measurement can put strong constraints on the models with gauge coupling unification. As already mentioned in \refsec{sec:mod}, it is expected that in the not-so-far future the present SK limit will be extended by at least one order of magnitude, 
\be\label{eq.prot_HK}
\tau_p\gsim 2\times 10^{35}\;\; \textrm{years},
\ee 
by a new generation detector Hyper-Kamiokande\cite{Abe:2018uyc}. 

\begin{figure}[t]
\centering
\subfloat[Testable by HK - model SS7$^{(5,6)}$]{%
\includegraphics[width=0.37\textwidth]{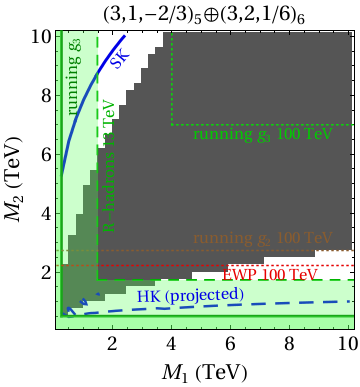}
}%
\hspace{0.3cm}
\subfloat[Testable by HK - model FS1$^{(18,6)}$]{%
\includegraphics[width=0.37\textwidth]{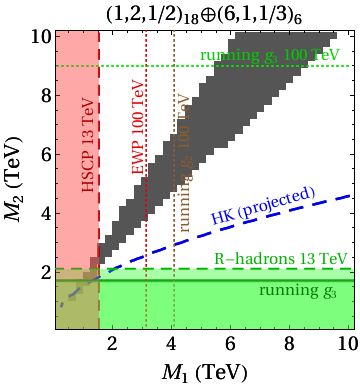}
}%
\caption{\footnotesize Summary of experimental bounds on masses $M_1$ and $M_2$ for scenarios (a) SS7$^{(5,6)}$ and (b) FS1$^{(18,6)}$. The color code is the same as in \reffig{fig:bounds1}. Blue solid line marks the exclusion by the proton decay measurement at Super-Kamiokande\cite{Miura:2016krn}. Blue dashed lines indicate the projected reach of Hyper-Kamiokande\cite{Abe:2018uyc}.
\label{fig:bounds3}}
\end{figure}

In column 6 of \reftables{2reps_FS}{2reps_SS} we report the maximal predicted proton lifetime in years, $\tau_p^{max}$, in the PGU region of the corresponding model. If this number is lower than the bound of \refeq{eq.prot_HK}, the model will be entirely tested by the HK measurement. This is exactly the case of scenarios FS1$^{(18,6)}$ and SS7$^{(5,6)}$ whose parameter space is presented in \reffig{fig:bounds3}. The projected exclusion by HK is indicated as blue dashed lines, while a blue solid line denotes the present day SK bound. 

The PGU scenarios in \reffig{fig:bounds3} are characterized by the spectra of type H3 and H0, correspondingly. Here both BSM particles can be very heavy and for this reason the PGU parameter space is difficult to probe with the collider searches. The proton lifetime measurement, however, can constrain scenarios with a relatively low unification scale independently on the BSM fermion and scalar masses, and thus can provide yet another complementary way of testing the parameter space consistent with the precise unification condition. 

\subsection{LHC searches for long-lived particles}\label{ssec:dep}

In this study we make an assumption that the Yukawa interactions of the BSM fermions and scalars with the SM quarks and leptons are negligibly small, so that they can be treated as very long-lived, or quasi-stable particles which do not decay inside the detector.\footnote{If a Yukawa coupling is tiny but non-negligible, the lifetime of a BSM particle can vary from picoseconds to tens of nanoseconds. We leave a detailed discussion of the experimental bounds on such feeble Yukawa couplings for future study.} Dedicated analyses looking for such heavy stable charged particles (HSCP) were performed both by the ATLAS and CMS collaborations, based on the observables related to ionization energy loss and time of flight, characteristic of heavy particles whose velocities are significantly smaller than the speed of light.

\paragraph{Color HSCPs}

It is usually assumed that if the lifetime of a color  HSCP is longer than the typical hadronization time scale, it would form a colorless QCD bound state with the SM quarks and gluons, the so-called R-hadron. 

The most recent 95\% confidence level (C.L.) upper bound on the color HSCP production cross section, which can be translated into a lower bound on the HSCP mass, were derived by ATLAS using a data sample corresponding to 36 fb$^{-1}$ of proton-proton collisions at $\sqrt{s}=13$ TeV\cite{Aaboud:2019trc}, and by CMS using 2.5 fb$^{-1}$ of data at the same energy\cite{Khachatryan:2016sfv}. While in both analyses the HSCP under study is gluino (a supersymmetric partner of the gluon), the cross-section bounds can be easily applied to any color HSCP. The reason is that hadronic interactions of R-hadrons with detector material proceed through its light-quark constituents, and thus are to a good approximation independent on the original HSCP quantum numbers. 

In hadron colliders a color particle can be pair produced at the leading order (LO) through gluon fusion and quark-antiquark annihilation. 
A detailed calculation of the corresponding $p\,p\to X\,\bar{X}$ cross section for a BSM particle $X$ in a generic representation $R_3$ of $SU(3)_C$ is presented in Appendix~\ref{app:cs}. Comparing those results with the observed exclusion limits on the cross section reported by ATLAS\cite{Aaboud:2019trc}, we can derive lower bounds on the masses of colored particles in our PGU models. Those bounds are shown in column 4 of \reftables{2reps_FS}{2reps_SS}.

In \reffigsd{fig:bounds1}{fig:bounds3} the bounds from the R-hadron searches are shown as green dashed lines, to emphasize their dependence on the negligible Yukawa coupling assumption. In the presence of non-zero Yukawa interactions, color BSM fermions and scalars would decay before a bound state is formed and the limits would not apply anymore. This is in striking contrast with the bounds from the running strong coupling constant, which are model-independent once the $SU(3)_C$ charges are fixed.

Note that at the moment the R-hadron limits are stronger (often significantly) than those from the running $g_3$, and so it is due to them that many of the PGU scenarios can be excluded. We indicate this fact with an asterisk in column 4 of \reftables{2reps_FS}{2reps_SS}.

\paragraph{EW-charged HSCPs}

If a HSCP is a $SU(3)_C$ singlet but it carries non-zero EW charges, it will be produced through Drell-Yan (DY) processes and will predominantly lose energy via ionization inside the detector. The corresponding LO cross section for a BSM particle with arbitrary $SU(2)_L$ and $U(1)_Y$ quantum numbers is calculated in Appendix~\ref{app:cs}. The lower bounds on the BSM particle masses can be derived using the chargino- and tau-dedicated  HSCP searches by ATLAS with 36 fb$^{-1}$ of data\cite{Aaboud:2019trc}. 

The corresponding exclusion bounds are summarized in column 5 of \reftables{2reps_FS}{2reps_SS} and depicted in \reffigsd{fig:bounds1}{fig:bounds3} as dashed red lines. At the moment they provide the only experimental way of testing at the colliders the PGU scenarios in which the non-color particle is lighter than the color one. In the future competing bounds, independent on values of the BSM Yukawa couplings, should be provided by the EW searches discussed in \refsec{sub:in}.

\subsection{Synopsis of the results}

In this section we performed a detailed analysis of various experimental signatures characteristic of the PGU scenarios identified in \refsec{sec:mod}.

We found that out of 67(78) models of type FS(SS) that allow for the precise gauge coupling unification, 23(35) are already excluded (up to 10 \tev) by the combination of the running strong coupling measurement and the HSCP searches. Those models are dubbed  as ``excluded'' in column 10 of \reftables{2reps_FS}{2reps_SS}. 

We have also shown that many of the remaining scenarios, 23 in the FS case and 25 in the SS case, could be further probed at the future  100\tev\ proton collider, or by the proton lifetime measurement at HK. Those models are dubbed  as ``testable'' in column 10 of \reftables{2reps_FS}{2reps_SS}. Note that they almost exclusively belong to the categories H1 and H2, with a few H0 and H3 exceptions that can be either tested by HK, or are border-line cases between different types of spectra.

Finally, there are 21(18) models of type FS(SS) that remain very challenging to probe experimentally. Those are predominantly of the H3 and H0 spectrum type, i.e.~featuring the parameter space where both BSM particles can be heavy. Moreover, in many of these models the unification scale is very high, $\mgut>10^{17}\gev$, so there is not much hope to probe them in HK-like experiments either. 

Nevertheless, it feels encouraging that the great majority of the PGU models identified in this study will be experimentally probe in the next decades.


\section{Conclusions}\label{sec:con}

In this study we performed a systematic analysis of the SM extensions with BSM scalars and fermions in different representations of the SM gauge group in the context of the gauge coupling unification. We analyzed all possible combinations of one representation of scalars and one representation of fermions, as well as of two different representations of scalars, where the number of fermions and scalars in each representation was limited by demanding perturbativity of the gauge couplings at the unification scale. 

We found 145 different two-representation BSM extensions that allow for precise gauge coupling unification at the energies higher than $10^{15}\gev$, under the assumption that all BSM masses are in the range $0.25-10\tev$. The successful models are listed in \reftables{2reps_FS}{2reps_SS} of Appendix D, together with the information about the type of spectrum favored by the unification condition. The latter is particularly important in the context of experimental testability of the unification models. In particular, those PGU scenarios which feature hierarchical spectra, with one of the BSM particles significantly lighter than the other, are very susceptible to collider searches based on heavy stable charged particle signatures, as well as to direct measurements of the SM gauge coupling scale dependence. 

We also found that complementarity of various experimental strategies plays a crucial role in probing the parameter space of the unification scenarios. Out of 145 possible models, 40\% can already be excluded based on the present data, while 33\% more may be probed in the future at the 100\tev\ collider. This observation is in line with the findings of\cite{Kowalska:2019qxm}, where VL fermion extensions of the SM were analyzed. 

We hope that the results of our study would add yet another small incentive in favor of building a more powerful proton collider.

\bigskip\bigskip
\noindent \textbf{ACKNOWLEDGMENTS}
\medskip

\noindent 
U.C.O. would like to thank Piotr Chankowski and Paweł Nurowski for helpful comments and discussions. This work is supported by the National Science Centre (Poland) under the research Grant No.~2017/26/E/ST2/00470. U.C.O is partially supported by the Polish National Agency for Academic Exchange (NAWA) via the Polish Returns 2019 program. The use of the CIS computer cluster at the National Centre for Nuclear Research in Warsaw is gratefully acknowledged.

\appendix

\section{Group invariants and loop coefficients}\label{app:rges}

Let us consider a symmetry group $G$. The quadratic Casimir operator for the representation $R$ is defined as
\be
 C(R)\delta^i_j=(t^At^A)^i_j=\sum_{A=1}^d t^At^A,
\ee
where $t^A$ are the generators of $G$ in the representation $R$. The Dynkin index of a representation $R$ is instead given by
\be
 S(R)\delta^{A\,B}=\textrm{Tr}\left\{t^At^B\right\}.
\ee
The two are related through the dimension of the adjoint representation $d(\textrm{Adj})$,
\be
 S(R)d(\textrm{Adj})=C(R)d(R).
\ee
For convenience, one can parametrize the group-theoretical factors through the weights $(p,q)$ for an irreducible $SU(3)$ representation $R_3$, and through the highest weight $\ell$ for an $SU(2)$ representation $R_2$, 
\be
\begin{array}{l}
d(R_3)=\frac12(p+1)(q+1)(p+q+2)\,,\\[1ex]
C(R_3)=p+q+\frac13 (p^2+q^2+p q)\,,\quad {\rm with}\quad p,q=0,1\cdots\,,\\[1ex]
d(R_2)=2\ell+1\,,\\[1ex]
C(R_2)=\ell(\ell+1) \,, \quad{\rm with}\quad  \ell=0,\frac{1}{2},1\cdots\,.
\end{array}
\ee

General two-loop beta functions for a system of gauge couplings $g_i$ of a direct-product symmetry group $G_i\times\cdots$ read\cite{Machacek:1983tz}
\bea
\label{eq:RGEtwo}
\beta_{i} =\frac{dg}{d\ln \mu}&=& \frac{g_i^3}{(4\pi)^2}\left[-\frac{11}{3}C(G_i)+\frac{2}{3}S_i(R_F)+\frac{1}{3}S_i(R_S)\right]\\
&+&\frac{g_i^5}{(4\pi)^4}\left[-\frac{34}{3} C(G_i)^2+\Big(2C_i(R_{F})+\frac{10}{3}C(G_i)\Big)S_i(R_F)+\Big(4C_i(R_S)+\frac{2}{3}C(G_i)\Big)S_i(R_S) \right.\nonumber\\
&+&\left.\sum_{j=1}^kg_j^2\Big(2C_j(R_F)S_i(R_F)+4C_j(R_S)S_i(R_S)\Big)\right],\nonumber
\eea
where $G_i$, $R_F$ and $R_S$ indicate contributions from gauge bosons, Weyl fermions, and complex scalars respectively. The sum is meant in both $S(R_F)$ and $S(R_S)$ over all fermion and scalar representations transforming nontrivially under $G_i$. 

The two-loop beta functions for SM augmented with $N_F$ Weyl fermions in the representation $(R_{F3},R_{F2},Y_F)$ and $N_S$ complex scalars in the representation $(R_{S3},R_{S2},Y_S)$ can be straightforwardly derived from \refeq{eq:RGEtwo}, and read
\bea
\label{beta1}\beta_3 &=& \frac{g_3^3}{(4\pi)^2} B_3+\frac{g_3^3}{(4\pi)^4}\left(C_{33}\, g_3^2  + C_{32}\, g_2^2 +C_{31}\, g_1^2\right),\\
\label{beta2}\beta_2 &=& \frac{g_2^3}{(4\pi)^2} B_2+\frac{g_2^3}{(4\pi)^4}\left(C_{23}\, g_3^2  + C_{22}\, g_2^2 +C_{21}\, g_1^2\right),\\
\label{beta3}\beta_1 &=& \frac{g_1^3}{(4\pi)^2} B_1+\frac{g_1^3}{(4\pi)^4}\left(C_{13}\, g_3^2  + C_{12}\, g_2^2 +C_{11}\, g_1^2\right),
\eea
with the one-loop coefficients defined as
\bea
B_3 &=& -7+\frac{2}{3}N_F\,S(R_{F3})\,d(R_{F2})+\frac{1}{3}N_S\,S(R_{S3})\,d(R_{S2}), \label{oneloop3}\\[1.5ex]
B_2 &=&-\frac{19}{6}+\frac{2}{3} N_F\, S(R_{F2})\,d(R_{F3})+\frac{1}{3} N_S\, S(R_{S2})\,d(R_{S3}), \label{oneloop2}\\[1.5ex]
B_1 &=&\frac{41}{10}+\frac{2}{5} N_F\, d(R_{F3}) d(R_{F2}) Y_F^2+\frac{1}{5} N_S\, d(R_{S3}) d(R_{S2}) Y_S^2.\label{oneloop1}
\eea
The two-loop coefficients are instead given by
\bea\label{twoloop}
C_{33}&=&-26+ N_F \,S(R_{F3})\,d(R_{F2})\big(2 C(R_{F3})+10\big)+ N_S \,S(R_{S3})\,d(R_{S2})\big(4 C(R_{F3})+2\big),\\[1.5ex]
C_{32}&=& \frac{9}{2}+ 2 N_F \,S(R_{F3}) \,C(R_{F2}) \,d(R_{F2})+ 4 N_S \,S(R_{S3}) \,C(R_{S2}) \,d(R_{S2}),\\[1.5ex]
C_{31}&=&\frac{11}{10} +\frac{6}{5} N_F\, S(R_{F3}) d(R_{F2}) Y_F^2+\frac{12}{5} N_S\, S(R_{S3}) d(R_{S2}) Y_S^2,\\[1.5ex]
C_{23}&=&12+2 N_F\,S(R_{F2})\, C(R_{F3})\, d(R_{F3})+4 N_S\,S(R_{S2})\, C(R_{S3})\, d(R_{S3}),  \\[1.5ex]
C_{22}&=&\frac{35}{6}+ N_F\, S(R_{F2}) \,d(R_{F3})\big( 2 C(R_{F2})+  \frac{20}{3}\big)+ N_S\, S(R_{S2}) \,d(R_{S3})\big(4 C(R_{S2})+  \frac{4}{3}\big),\\[1.5ex]
C_{21}&=&\frac{9}{10} +\frac{6}{5}  N_F\, S(R_{F2}) d(R_{F3}) Y_F^2+\frac{12}{5}  N_S\, S(R_{S2}) d(R_{S3}) Y_S^2, \\[1.5ex]
C_{13}&=& \frac{44}{4} +\frac{6}{5} N_F\, C(R_{F3}) d(R_{F3})  d(R_{F2})  Y_F^2+\frac{12}{5}N_S\, C(R_{S3}) d(R_{S3})  d(R_{S2})  Y_S^2, \\[1.5ex]
C_{12}&=& \frac{27}{10} +\frac{6}{5} N_F\, C(R_{F2}) d(R_{F2}) d(R_{F3}) Y_F^2+\frac{12}{5} N_F\, C(R_{F2}) d(R_{F2}) d(R_{F3}) Y_F^2, \\[1.5ex]
C_{11}&=&\frac{199}{50} + \frac{18}{25} N_F\, d(R_{F3}) d(R_{F2}) Y_F^4+ \frac{36}{25} N_S\, d(R_{S3}) d(R_{S2}) Y_S^4.
\eea

\section{Decomposition of the irreducible $SU(5)$ representations}\label{su5}

In this Appendix we collected the branching rules for the embedding $SU(5)\supset SU(3)\times SU(2)\times U(1)$\cite{SLANSKY19811}.

\bea\label{su5reps}
{\bf 5}&= & \left({\bf 1}, {\bf 2}, \textstyle\frac{1}{2}\right)\oplus \left({\bf 3}, {\bf 1}, -\textstyle\frac{1}{3}\right),\\
{\bf 10}&= & \left({\bf 1}, {\bf 1}, 1\right)\oplus \left({\bf\bar{3}}, {\bf 1}, -\textstyle\frac{2}{3}\right)\oplus \left({\bf 3}, {\bf 2}, \textstyle\frac{1}{6}\right),\nonumber\\
{\bf 15}&= & \left({\bf 1}, {\bf 3}, 1\right)\oplus \left({\bf 3}, {\bf 2}, \textstyle\frac{1}{6}\right)\oplus \left({\bf 6}, {\bf 1}, -\textstyle\frac{2}{3}\right),\nonumber\\
{\bf 24}&= & \left({\bf 1}, {\bf 1}, 0\right)\oplus \left({\bf 1}, {\bf 3}, 0\right)\oplus \left({\bf 8}, {\bf 1}, 0\right)\oplus\left({\bf 3}, {\bf 2}, -\textstyle\frac{5}{6}\right)\oplus\left({\bf\bar 3} , {\bf 2}, \textstyle\frac{5}{6}\right),\nonumber\\
{\bf 35}&= & \left({\bf 1}, {\bf 4}, -\textstyle\frac{3}{2}\right)\oplus \left({\bf\bar 3}, {\bf 3}, -\textstyle\frac{2}{3}\right)\oplus \left({\bf\bar 6}, {\bf 2}, \textstyle\frac{1}{6}\right)\oplus \left({\bf\bar 10}, {\bf 1}, 1\right),\nonumber\\
{\bf 40}&= & \left({\bf 1}, {\bf 2}, -\textstyle\frac{3}{2}\right)\oplus \left({\bf 3}, {\bf 2}, \textstyle\frac{1}{6}\right)\oplus \left({\bf\bar 3}, {\bf 1}, -\textstyle\frac{2}{3}\right)\oplus \left({\bf\bar 3}, {\bf 3}, -\textstyle\frac{2}{3}\right)\oplus \left({\bf 8}, {\bf 1}, 1\right)\oplus \left({\bf\bar 6}, {\bf 2}, \textstyle\frac{1}{6}\right),\nonumber\\
{\bf 45}&=& \left({\bf 1}, {\bf 2}, \textstyle\frac{1}{2}\right)\oplus \left({\bf 3}, {\bf 1}, -\textstyle\frac{1}{3}\right)\oplus \left({\bf 3}, {\bf 3}, -\textstyle\frac{1}{3}\right)\oplus \left({\bf\bar 3}, {\bf 1}, \textstyle\frac{4}{3}\right)\oplus \left({\bf\bar 3}, {\bf 2}, -\textstyle\frac{7}{6}\right)\oplus \left({\bf\bar 6}, {\bf 1}, -\textstyle\frac{1}{3}\right)\oplus \left({\bf 8}, {\bf 2}, \textstyle\frac{1}{2}\right),\nonumber\\
{\bf 50}&= & \left({\bf 1}, {\bf 1}, -2\right)\oplus \left({\bf 3}, {\bf 1}, -\textstyle\frac{1}{3}\right)\oplus \left({\bf\bar 3}, {\bf 2}, -\textstyle\frac{7}{6}\right)\oplus \left({\bf\bar 6}, {\bf 3}, -\textstyle\frac{1}{3}\right)\oplus \left({\bf 6}, {\bf 1}, \textstyle\frac{4}{3}\right)\oplus \left({\bf 8}, {\bf 2}, \textstyle\frac{1}{2}\right),\nonumber\\
{\bf 70}&= & \left({\bf 1}, {\bf 2}, \textstyle\frac{1}{2}\right)\oplus \left({\bf 1}, {\bf 4}, \textstyle\frac{1}{2}\right)\oplus \left({\bf 3}, {\bf 1}, -\textstyle\frac{1}{3}\right)\oplus \left({\bf 3}, {\bf 3}, -\textstyle\frac{1}{3}\right)\oplus \left({\bf\bar 3}, {\bf 3}, \textstyle\frac{4}{3}\right)\oplus \left({\bf 6}, {\bf 2}, -\textstyle\frac{7}{6}\right)\oplus \left({\bf 8}, {\bf 2}, \textstyle\frac{1}{2}\right)\oplus \left({\bf 15}, {\bf 1}, -\textstyle\frac{1}{3}\right),\nonumber\\
{\bf 70^{\prime}}&= &\left({\bf 1}, {\bf 5}, -2\right)\oplus \left({\bf\bar 3}, {\bf 4}, -\textstyle\frac{7}{6}\right)\oplus \left({\bf\bar 6}, {\bf 3}, -\textstyle\frac{1}{3}\right)\oplus \left({\bf \overline{10}}, {\bf 2}, \textstyle\frac{1}{2}\right)\oplus \left({\bf\overline{15}^{\prime}}, {\bf 1}, \textstyle\frac{4}{3}\right),\nonumber\\
{\bf 75^{\prime}}&= &\left({\bf 1}, {\bf 1}, 0\right)\oplus \left({\bf 3}, {\bf 1}, \textstyle\frac{5}{3}\right)\oplus \left({\bf 3}, {\bf 2}, -\textstyle\frac{5}{6}\right)\oplus \left({\bf\bar 3}, {\bf 1}, -\textstyle\frac{5}{3}\right)\oplus \left({\bf\bar 3}, {\bf 2}, \textstyle\frac{5}{6}\right)\oplus \left({\bf\bar 6}, {\bf 2}, -\textstyle\frac{5}{6}\right)\oplus \left({\bf 6}, {\bf 2}, \textstyle\frac{5}{6}\right)\oplus\nonumber\\
&&\left({\bf 8}, {\bf 1}, 0\right)\oplus \left({\bf 8}, {\bf 3}, 0\right).\nonumber
\eea

\section{Production cross section of the BSM fermions and scalars at the LHC}\label{app:cs}

In this section we derive analytical formulae for the pair-production cross section of color and non-color fermions and scalars at the LHC. We further confront our results with the output of numerical calculations performed with \madgr\cite{Alwall:2014hca}.

\subsection{Color particles}\label{app:cscol}

Pair production of heavy color particles at hadron colliders proceeds dominantly via strong interactions. At the LO, two classes of processes contribute to the parton-level production cross section: quark-antiquark annihilation and gluon fusion. The numerical value of the cross section is then entirely determined by three properties of the produced particle: its spin, its mass, and its transformation properties under the $SU(3)_C$ symmetry group, encoded in the representation $R_3$. 

\paragraph{Scalars}

\begin{figure}[t]
	\centering
	\includegraphics[width=0.21\linewidth]{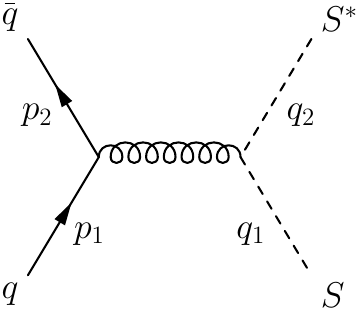}
	\caption{Feynman diagram for the LO scalar pair production at hadron colliders via quark-antiquark annihilation.}
	\label{fig:scalar_qq}
\end{figure}

The only process contributing to the quark-antiquark production of a color complex scalar $S$, $q\,\bar{q}\to S\,S^\ast$, is depicted in \reffig{fig:scalar_qq}, and its partonic cross section reads 
\be\label{eq_csSC1}
	\hat{\sigma} (q\,\bar{q}  \to S\,S^\ast) = \frac{d(\textrm{Adj})\,S(\textrm{Fun})S(R_3)}{d(\textrm{Fun})^2}\frac{\pi\alpha_{s}^{2}}{3\,\hat{s}}\beta_S^{3},
\ee
where $\beta_S = \sqrt{1-4m_{S}^2/\hat{s}}$, $\alpha_s=g_3^2/(4\pi)$, $m_S$ is the mass of the scalar $S$, the Mandelstam variable reads $\hat{s} = (p_{1} + p_{2})^{2} =  (q_{1} + q_{2})^{2}$, and $p_{i}$, $q_{i}$ denote the momenta of the initial and final state particles, respectively. ``Fun'' and ``Adj'' stand for the fundamental and the adjoint representation of $SU(3)_C$, correspondingly.

\begin{figure}[t]
\centering
\includegraphics[width=0.21\linewidth]{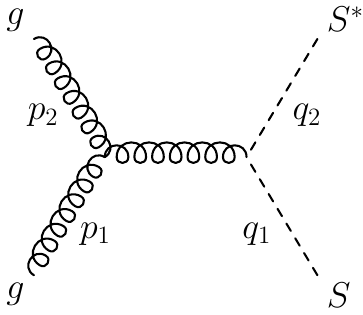}
\hfil
\includegraphics[width=0.21\linewidth]{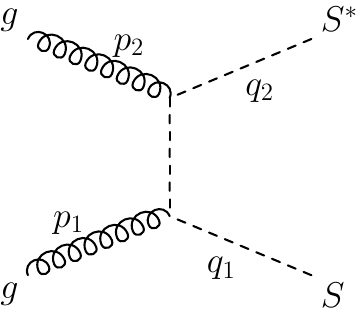}
\hfil
\includegraphics[width=0.21\linewidth]{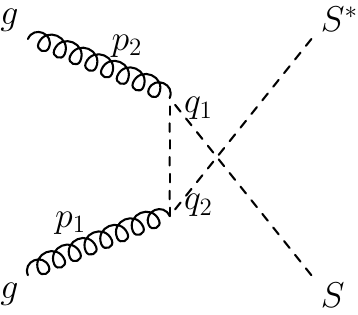}
\hfil
\includegraphics[width=0.19\linewidth]{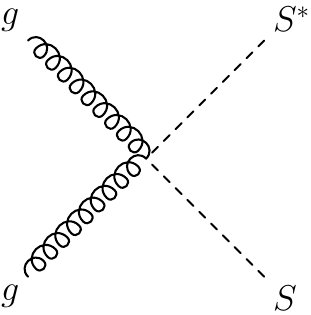}
\caption{Feynman diagrams for the LO scalar pair-production at hadron colliders via gluon fusion.}
\label{fig:scalar_gg}
\end{figure}
	
On the other hand, a contribution to the scalar pair production stemming from the gluon fusion, $g\,g\to S\,S^\ast$, is less straightforward to calculate. The LO Feynman diagrams are depicted in \reffig{fig:scalar_gg} and the corresponding amplitudes read, from left to right,
\bea\label{eq:matrix_element_gg_SS}
i\mathcal{M}_{s} &=& -g_3^{2}\,f^{abc}\, (T^{c})_{ij}\frac{\widetilde{\mathcal{M}}_s}{\hat{s}},\label{eq:ggSS_s}\\
i\mathcal{M}_{t} &=& ig_3^{2}\,(T^{a}T^{b})_{ij} \frac{\widetilde{\mathcal{M}}_t}{t-m_{S}^{2}},\\
i\mathcal{M}_{u} &=& ig_3^{2}\,(T^{b}T^{a})_{ij}\frac{\widetilde{\mathcal{M}}_u}{u-m_{S}^{2}},\\
i\mathcal{M}_{tu} &=& ig_3^{2}\, \{T^{a},T^{b}\}_{ij}\,\widetilde{\mathcal{M}}_{tu}.\label{eq:ggSS_tu}
\eea
In the above $T^{a}$ are the generators of $SU(3)$, $f^{abc}$ are the corresponding structure constants, and $t = (p_{1}-q_1)^{2}$ and $u=(p_{1}-q_2)^{2}$ are the Mandelstam variables. The objects $\widetilde{\mathcal{M}}_{i}$ capture the kinematics of the process and are given by
\bea
\widetilde{\mathcal{M}}_{s} 
		&=& \left[\epsilon(p_{2})\cdot(q_{1} - q_{2})\right]\,\left[\epsilon(p_{1})\cdot(2\,p_{2}+p_{1})\right]-\left[\epsilon(p_{2})\cdot (2\,p_{1}+p_{2})\right]\,\left[\epsilon(p_{1})\cdot(q_{1} - q_{2})\right]\label{eq:ggSS_Ms}	\\
				&+&\left[\epsilon(p_{1})\cdot \epsilon(p_{2})\right]\,\left[(p_{1}-p_{2})\cdot(q_{1} - q_{2})\right],\nonumber\\
\widetilde{\mathcal{M}}_{\,t}  
		&=& \left[\epsilon(p_{1})\cdot(2\,q_{1} - p_{1}) \right]\,\left[\epsilon(p_{2})\cdot(p_{2}-2\,q_{2})\right],\\
\widetilde{\mathcal{M}}_{\,u}  
		&=& \left[ \epsilon(p_{2})\cdot(p_{1}-2\,q_{2})\right] \,\left[\epsilon(p_{1})\cdot(2\,q_{1}-p_{2})\right],\\
\widetilde{\mathcal{M}}_{\,tu}
		&=& \epsilon(p_{1})\cdot \epsilon(p_{2}),\label{eq:ggSS_Mtu}
\eea
where $\epsilon$ is a polarization vector of the gluon and the contraction of the Lorentz indices is understood.

The calculation of the total $g\,g\to S\,S^\ast$ amplitude can be simplified by employing the factorization property of the non-Abelian gauge theories, according to which the amplitude of the process can be expressed as a product of two terms,
\be\label{eq:fact}
\mathcal{M}=\sum_{\alpha=s,t,u,tu}\mathcal{M}_\alpha=G\,\mathcal{M}_A,
\ee
where $G$ encodes the properties of the produced particle with respect to the non-Abelian gauge group, and $\mathcal{M}_A$ involves the dynamical part corresponding to the pure Abelian process. To derive the explicit forms of both factors, let us first note that the total $g\,g\to S\,S^\ast$ amplitude can be written as\cite{Zhu:1980sz,Blumlein:1996qp}
\be\label{eq:tot_amp}
\mathcal{M} = \sum_{\alpha=s,t,u}g_3^{2}\, G_{\alpha}\,\frac{T_{\alpha}}{C_{\alpha}},
\ee
where  $G_{s}=if^{abc}(T^{c})_{ij}$, $G_{t}=(T^{a}T^{b})_{ij}$, $G_{u}=(T^{b}T^{a})_{ij}$ are the non-Abelian group factors, and $C_{s}\equiv \hat{s}$, $C_{t}\equiv \hat{t}=t - m^{2}_{S}$, $C_{u}\equiv \hat{u}=u - m^{2}_{S}$ are the denominators in~\refeqst{eq:ggSS_s}{eq:ggSS_tu}.  The numerators $T_{\alpha}$ for each channel are straightforwardly obtained from~\refeqst{eq:ggSS_Ms}{eq:ggSS_Mtu} as
\bea\label{eq:lorentz_scalar_amp}
T_{s} &= & \widetilde{\mathcal{M}}_{\,s},\\
T_{t} &= &  \widetilde{\mathcal{M}}_{\,t} - \hat{t}\,\widetilde{\mathcal{M}}_{\,tu},\\
T_{u} &= & \widetilde{\mathcal{M}}_{\,u} - \hat{u}\,\widetilde{\mathcal{M}}_{\,tu}.
\eea
It is now easy to verify that the factors appearing in \refeq{eq:tot_amp} satisfy the following relations,
	\bea\label{eq:CGT}
			C_{s} + C_{t} + C_{u} =  0,\\
			G_{t} - G_{u} -G_{s} = 0, \nonumber\\
			T_{t} - T_{u} -T_{s} =0 \nonumber.
	\eea
From \refeq{eq:tot_amp} and \refeq{eq:CGT}  it results that the group theoretical factor $G$ and the Abelian amplitude $\mathcal{M}_A$ in \refeq{eq:fact} read
\be
G =\frac{C_{u}G_{t} + C_{t}G_{u}}{-C_{s}},\qquad \mathcal{M}_A=g_3^{4}\left( \frac{T_{t}}{C_{t}} + \frac{T_{u}}{C_{u}}\right).
\ee
Summing over the final and averaging over the initial degrees of freedom (colors and spins), one obtains the averaged squared amplitude
\be\label{eq:squaredmodulus}
\overline{	\left|\mathcal{M} \right|^{2}}= {\overline{\left|G \right|^{2}}} \cdot {	\overline{\left|\mathcal{M}_A\right|^{2}}},
\ee
where
\be
\overline{\left|G \right|^{2}} = \frac{1}{d(\text{Adj})^{2}}d(R_3)\,C(R_3)\left(C(R_3) -C(\textrm{Adj})\frac{\hat{u}\,\hat{t}}{\hat{s}^{2}}\right),
\ee
and
\be
\overline{\left|\mathcal{M}_A\right|^{2}} = 2\,g_3^4\left[1-2\frac{\hat{s}}{\hat{u}\,\hat{t}}\,m_{S}^{2} +2\left(\frac{\hat{s}}{\hat{u}\,\hat{t}}\,m_{S}^{2}\right)^{2}\,\right].
\ee
Finally, the partonic cross section for the process $g\,g\to S\,S^\ast$ can be written as
\bea\label{eq_csSC2}
\hat{\sigma} (g\,g \to S\,S^\ast) &=&\frac{d(R_3)\,C(R_3)}{d(\text{Adj})^{2}} \frac{ \pi	\alpha_{s}^{2}}{6\hat{s}}\Bigg[C(\text{Adj})\beta_S(3-5\beta_S^{2}) - 12C(R_3)\beta_S(\beta_S^{2}-2)\\ 
			&&\qquad\qquad\qquad\;\;+\left. \text{ln}\left|\frac{\beta_S +1}{\beta_S -1}\right|\left[6C(R_3)(\beta_S^{4}-1)-3C(\text{Adj})(\beta_S^{2}-1)^{2}\right]\right].\nonumber
\eea
The above formula reproduces the results of\cite{Blumlein:1996qp,PhysRevD.79.054002} for color triples and sextets, correspondingly.

\paragraph{Fermions}

The derivation of the total partonic cross sections for the processes $q\,\bar{q}  \to F\bar{F}$ and $g \,g \to F\bar{F}$ proceeds analogously to the scalar case. The corresponding Feynman diagrams can be straightforwardly obtained from those depicted in \reffigs{fig:scalar_qq}{fig:scalar_gg} by replacing the scalar external lines with the fermion ones (the only exception is the rightmost diagram in \reffig{fig:scalar_gg}, which does not exists for the fermions). The factorization conditions \refeq{eq:fact} and \refeq{eq:CGT} are satisfied for the fermion case as well, although the kinematical factors $T_{\alpha}$ assume a slightly different form. In the end, 	the total partonic cross sections read
\bea\label{eq:cs_ferC}
\hat{\sigma}_{q\,\bar{q}  \to F\bar{F}}(\hat{s})& =&  \frac{d(\textrm{Adj})\,S(\textrm{Fun})S(R_3)}{d(\textrm{Fun})^2}\frac{2\,\pi\alpha_{s}^{2}}{3\, \hat{s}}\beta_F\left(3-\beta_F^{2}\right),\\
\hat{\sigma}_{g \,g \to F\bar{F}}(\hat{s})& =& \frac{d(R_3)C(R_3)}{d(\text{Adj})^{2}} \frac{\pi  \alpha_{s}^{2}}{4\hat{s}}\Bigg[\frac{4}{3}C(\text{Adj})\beta_F+2\beta_F(2-\beta_F^{2} )( C(R_3)-\frac{22}{3}C(\text{Adj}))\nonumber  \\
			&& \qquad\qquad\qquad + \text{ln}\left|\frac{\beta_F +1}{\beta_F -1}\right|\left[8C(\text{Adj})(2-\beta_F^2)-C(R_3)(3-\beta_F^4)\right]\Bigg],
\eea
with $\beta_F = \sqrt{1-4m_{F}^2/\hat{s}}$. \refeq{eq:cs_ferC} reproduces the results of Ref.\cite{PhysRevD.17.2324} for color triplets. 
	
\paragraph{Hadronic cross section}

\begin{figure}[t]
	\centering
	\includegraphics[width=0.4\linewidth]{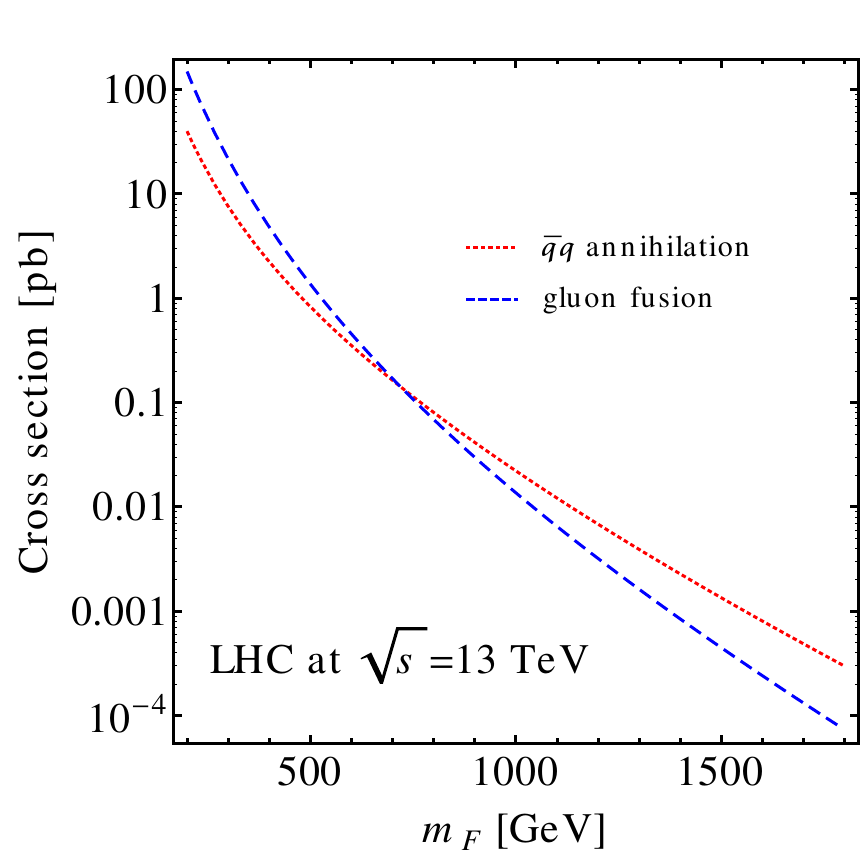}
	\hspace{0.05\textwidth}
	\includegraphics[width=0.41\linewidth]{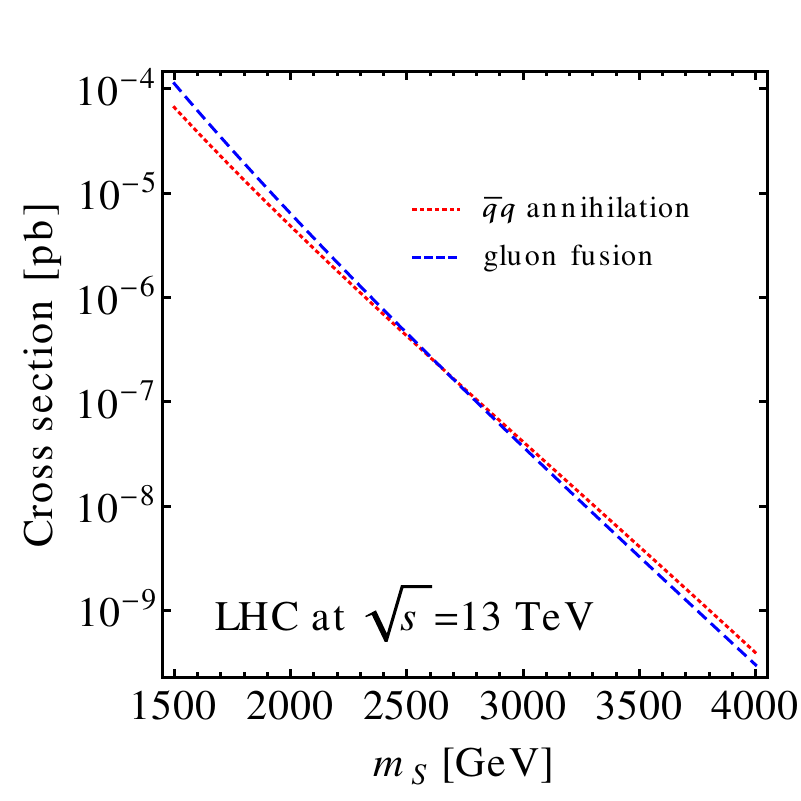}
	\caption{Contributions to the pair-production cross section at the LHC with $\sqrt{s}=13$ TeV of the $SU(3)_C$ triplet fermions (left) and scalars (right) as a function of their mass, stemming from the quark-antiquark annihilation (red dotted) and the gluon fusion (blue dashed).}
	\label{fig:cross_sec_channels}
\end{figure}	

The hadronic cross sections $\sigma (p\,p\to S\,S^\ast)$ and $\sigma (p\,p \to F\,\bar{F})$ are obtained by convoluting the corresponding partonic cross sections with the parton luminosties $L_{ab}^{AB}$ as
\be\label{eq:hadcs}
\sigma (p\, p\to X\bar{X}) = \int_{\tau_{0}}^{1} d\tau\left[\frac{d{L}_{gg}^{pp}}{d\tau}\hat{\sigma}_{gg\to X\bar{X}}(\hat{s}) + \frac{d{L}^{pp}_{q\bar{q}}}{d\tau}\hat{\sigma}_{q\bar{q}\to X\bar{X}}(\hat{s})\right],
\ee
with		
\be			
\frac{d{L}_{ab}^{AB}}{d\tau}= \frac{1}{1+\delta_{ab}}\int_{\tau}^{1}\frac{dx}{x}\left[f_{a}^{A}(x,\mu)f_{b}^{B}\left(\frac{\tau}{x},\mu\right)+f_{b}^{A}\left(\frac{\tau}{x},\mu\right)f_{a}^{B}\left(x,\mu\right)\right],
\ee
where $\tau=\hat{s}/s$, $\tau_{0}=4m^{2}_{S}/s$, $s$ is the hadronic center of mass energy squared and $f_{a}^{A}(x,\mu)$ indicates a parton distribution function for the parton $a$ in the hadron $A$ with the factorization scale $\mu$.	

\begin{figure}[t]
\centering
\includegraphics[width=0.4\linewidth]{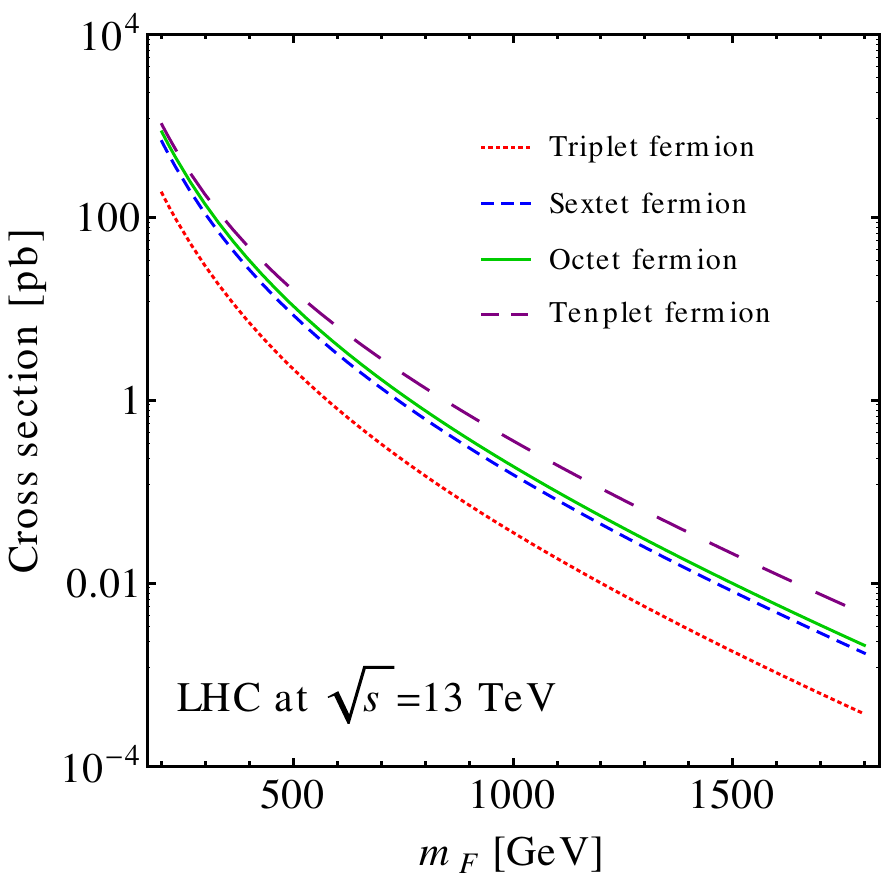}
\hspace{0.05\textwidth}
\includegraphics[width=0.4\linewidth]{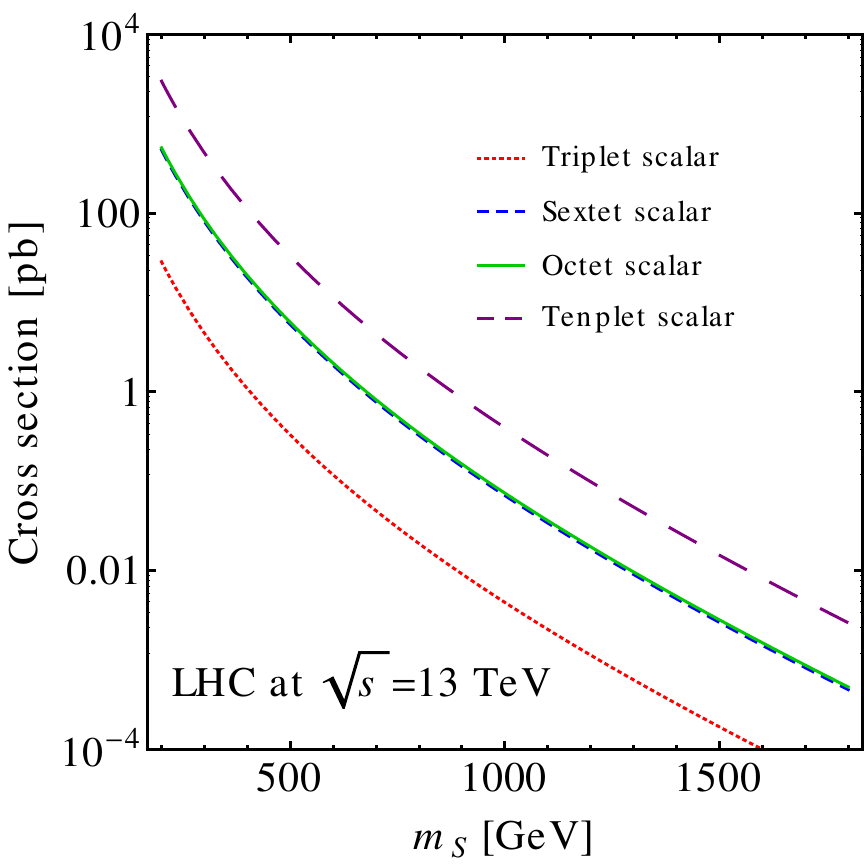}
\caption{Total hadronic cross section in $pp$ collisions at $\sqrt{s}=13$ TeV for the pair production of fermions (left panel) and scalars (right panel) in different $SU(3)_C$ representations: triplet (dotted red), sextet (dashed blue), octet (solid green) and tenplet (long-dashed purple).}
\label{fig:cross_sec_reps}
\end{figure}

In \reffig{fig:cross_sec_channels} we show the hadronic cross section at the LHC with $\sqrt{s}=13$ TeV  for the pair production of fermions (left panel) and scalars (right panel) as a function of their mass. We assume in both cases that the produced particle transforms as a $SU(3)_C$ triplet. Red dotted curve corresponds to the production via quark-antiquark annihilation, while the blue dashed one indicates the gluon fusion. One observes that for fermions lighter than $\sim 600\gev$ the gluon fusion is a dominant contribution to the total cross section, while for the heavier ones the quark-antiquark annihilation becomes larger and the cross section scales as $S(R_3)$. A similar pattern is observed in the scalar case, although the corresponding cross-over between the dominant contributions takes place for the scalar mass of around $\sim 2.5\tev$.

In \reffig{fig:cross_sec_reps} we present a comparison of the total hadronic cross sections for the pair production of fermions (left panel) and scalars (right panel) in different $SU(3)_C$ representations: triplet (dotted red), sextet (dashed blue), octet (solid green) and tenplet (long-dashed purple). The most important observation, which agrees with the findings of\cite{PhysRevD.79.054002}, is that the cross sections for sextets and octets are almost the same, as a direct consequence of similar numerical values of the group theoretical factors $C(R_3)$ and $S(R_3)$. 

\begin{figure}[t]
\centering
\includegraphics[width=0.40\linewidth]{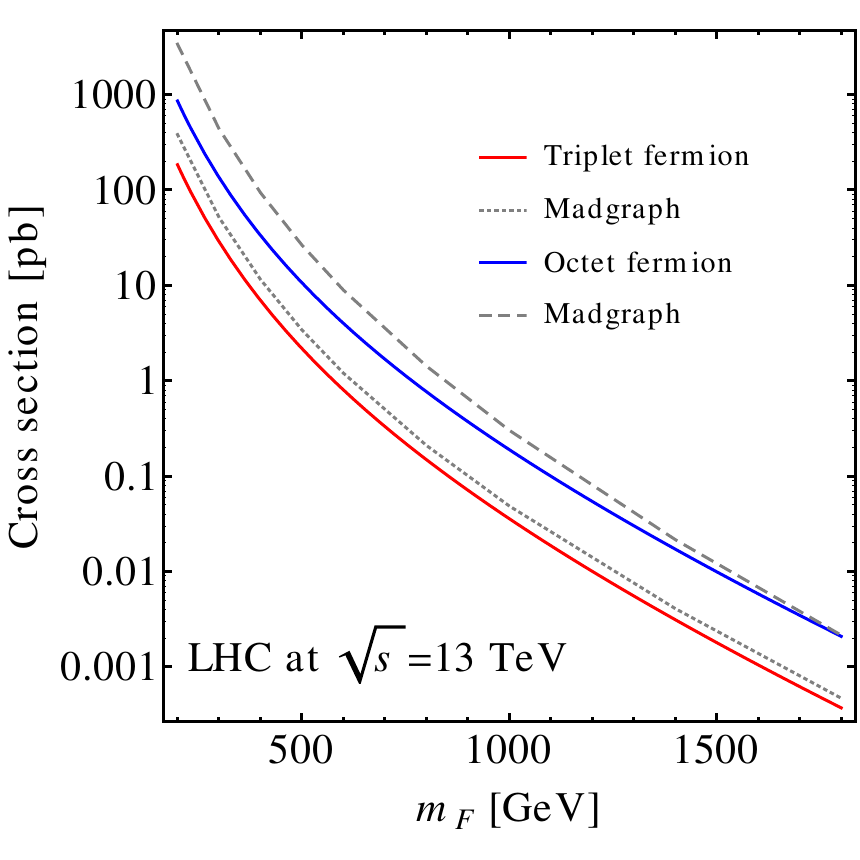}
\hspace{0.05\textwidth}
\includegraphics[width=0.4\linewidth]{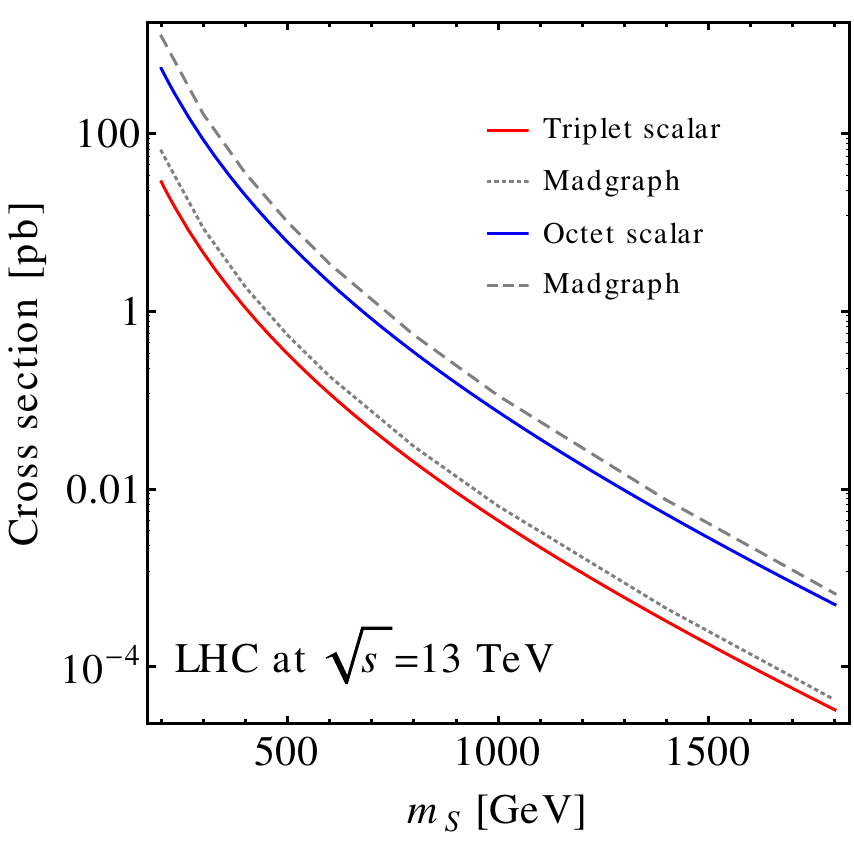}
\caption{Comparison of the analytical formulae \refeq{eq_csSC1} and \refeqst{eq_csSC2}{eq:hadcs} with the results of the numerical simulation performed with \madgr, for fermions (left panel) and scalars (right panel). $SU(3)_C$ triplets are indicated as solid red lines, while the $SU(3)_C$ octets as solid blue lines. Dotted and dashed grey lines denote the \madgr\ results, correspondingly.}
\label{fig:cross_sec_Mad}
\end{figure}

Finally, in \reffig{fig:cross_sec_Mad} we present a comparison of our analytical derivation of the tree-level hadronic sections based on \refeq{eq_csSC1} and \refeqst{eq_csSC2}{eq:hadcs}, corrected with the k-factor of 1.5\cite{Tanabashi:2018oca} 
to account for higher-order QCD corrections, with the results of the numerical simulation performed with \madgr. Once more, the left panel illustrates the case of the fermions and the right one of the scalars. $SU(3)_C$ triplets are indicated as solid red lines, while the $SU(3)_C$ octets as solid blue lines. Dotted and dashed grey lines denote the \madgr\ results, correspondingly. One can see that our results reproduce the numerical simulation quite precisely, in particular at the heavier end of the spectrum.

\subsection{EW-charged particles}

\begin{figure}[b]
	\centering
	\includegraphics[width=0.21\linewidth]{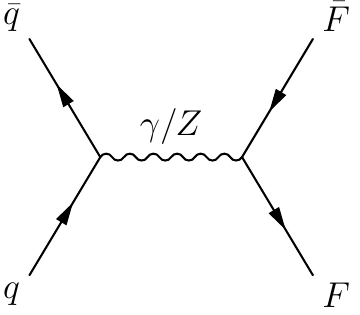}
	\hspace{1cm}
	\includegraphics[width=0.21\linewidth]{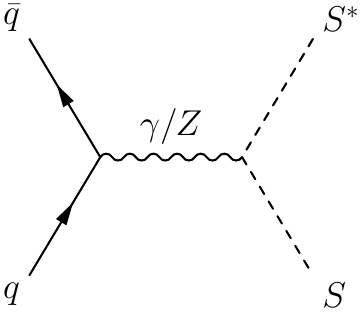}
	\caption{Feynman diagram for the LO fermion and scalar pair production at hadron colliders via the Drell-Yan process.}
	\label{fig:bsm_DY}
\end{figure}

At the tree -level, pair production of heavy non-color particles at hadron colliders proceeds through the Drell-Yan processes depicted in \reffig{fig:bsm_DY}. The corresponding amplitude can then be written as 
\be
\mathcal{M}=\mathcal{M}_\gamma+\mathcal{M}_Z.
\ee
Averaging over the initial color and spin states, and summing over the final spin states, one finds
\be\label{matrix_element_gamma_Z}
\overline{|\mathcal{M}|^2} =\frac{1}{3}\cdot \frac{1}{4} \left(\overline{|{\mathcal{M}_{\gamma}}|^{2}} + \overline{|\mathcal{M}_{Z}|^{2}} + 2\,\text{Re}\left(\overline{\mathcal{M}_{\gamma}\mathcal{M}_{Z}^{\dagger}}\right)\right).
\ee

In the following we present individual averaged squared contributions to \refeq{matrix_element_gamma_Z} for the scalars and fermions separately.

\paragraph{Fermions}

After the EWSB, the neutral current interactions of a generic fermion $F$ with the photon $A$ and the gauge boson $Z$ are given by
\be
\mathcal{L}_{\text{int}} =e\,Q_F\bar{F}\gamma^{\mu}A_\mu F +\frac{g_2}{\cos\theta_W}\bar{F}\gamma^{\mu}(c^F_{L}P_{L} + c^F_{R}P_{R})Z_{\mu}F,
\ee
where couplings of the fermion with $Z$ are entirely determined by its $SU(2)_L$ and $U(1)_Y$ quantum numbers as
\bea
c^F_{L/R}=\left(T_{3F}^{L/R}\cos^2\theta_W - Y_F^{L/R}\sin^2\theta_{W}\right).
\eea
Here $T_{3F}$ denotes the third component of the weak isospin, $\theta_W$ is the Weinberg angle, and the electric charge
$Q_F=T_{3F}+Y_F$.

The first term in \refeq{matrix_element_gamma_Z} resembles the matrix element for the process $q\,\bar{q}\to F\,\bar{F}$ with the gluon exchange in the $s$-channel. It reads
\be
\overline{|\mathcal{M}_{\gamma}|^{2}} = \frac{16(\pi\alpha\, Q_{q}Q_{F})^{2}}{3}\left[1+\beta^2_F\cos^2\theta+ \frac{4\,m_{F}^2}{\hat{s}}\right],
\ee
where $\alpha=e^2/(4 \pi)$ and $Q_q$ indicates the charge of the SM quark. The $s$-channel exchange of the boson $Z$ leads to
\bea
\overline{|\mathcal{M}_{Z}}|^{2}& = & \frac{4(\hat{s}\pi\alpha)^{2}}{3\cos^{4}\theta_{W}\sin^{4}\theta_{W}{\left[{(\hat{s}- M_{Z}^{2})^2} + (\Gamma_Z M_{Z})^2\right]}}\Bigg\{\frac{8\,m_{F}^{2}}{\hat{s}}c_{L}^{F}c_{R}^{F}\left[(c_{L}^{q})^{2} + (c_{R}^{q})^{2}\right]\\
&&+\left[(c_{L}^{q}c_{L}^{F})^{2} + (c_{R}^{q}c_{R}^{F})^{2}\right](1+\beta_{F}\cos\theta)^2
	+\left[(c_{L}^{q}c_{R}^{F})^{2} + (c_{R}^{q}c_{L}^{F})^{2}\right](1-\beta_{F}\cos\theta)^2\Bigg\},\nonumber
\eea
where $M_Z$ denotes the mass of the $Z$ boson, and $\Gamma_Z$ its total decay width. Finally, the interference term is given by
\bea
2\,\text{Re}\left(\overline{\mathcal{M}_{\gamma}\mathcal{M}_{Z}^{\dagger}}\right)&=&
\frac{8\hat{s}\pi^{2}\alpha^2 Q_{q}Q_{F}}{3\cos^2\theta_{W}\sin^2\theta_{W}} \times\frac{\hat{s}-M_{Z}^{2}}{(\hat{s}-M_{Z}^{2})^{2}+(\Gamma_Z M_{Z})^{2}}\\
&&\Bigg\{(c_{L}^{q}c_{R}^{F} + c_{R}^{q}c_{L}^{F})\left[(1-\beta_{F}\cos\theta)^{2} +\frac{4\,m_{F}^{2}}{\hat{s}}\right]+(c_{L}^{q}c_{L}^{F} + c_{R}^{q}c_{R}^{F})\left[(1+\beta_{F}\cos\theta)^{2}+\frac{4\,m_{F}^{2}}{\hat{s}}\right]\Bigg\}.\nonumber
\eea
Consequently, the partonic cross section for the pair production of colorless fermions reads
\bea
\hat{\sigma}_{q\,\bar{q}\to F\,\bar{F}}(\hat{s}) &=&\pi\alpha^2\beta_{F}\left\lbrace \frac{2 Q_{q}^2Q_{F}^{2}}{9 \hat{s}}\left(3-\beta_{F}^{2}\right) \right.\\
			& +& \frac{\hat{s}[(c_{L}^{q})^{2} + (c_{R}^{q})^{2}]}{36\cos^{4}\theta_{W}\sin^{4}\theta_{W}{\left[{(\hat{s}- M_{Z}^{2})^2} + (\Gamma_Z M_{Z})^2\right]}}\left[\left(3-\beta_{F}^{2}\right)\left(c_{L}^{F} + c_{R}^{F}\right)^{2} + 2\beta_{F}^{2}\left(c_{L}^{F} - c_{R}^{F}\right)^{2}\right]\nonumber\\
			& +&\left.\frac{Q_{q}Q_{F}}{9\cos^2\theta_{W}\sin^2\theta_{W}}\frac{(\hat{s}-M_{Z}^{2})(c_{L}^{q} + c_{R}^{q})(c_{L}^{F} +c_{R}^{F}) }{(\hat{s}-M_{Z}^{2})^{2}+(\Gamma_Z M_{Z})^{2}}\left(3-\beta_{F}^{2}\right)\right\rbrace.\nonumber
\eea
The above formula reproduces the results of\cite{Frampton:1992ik} for the $SU(2)_L$ singlets with an electric charge $1$.

\begin{figure}[t]
\centering
\includegraphics[width=0.40\linewidth]{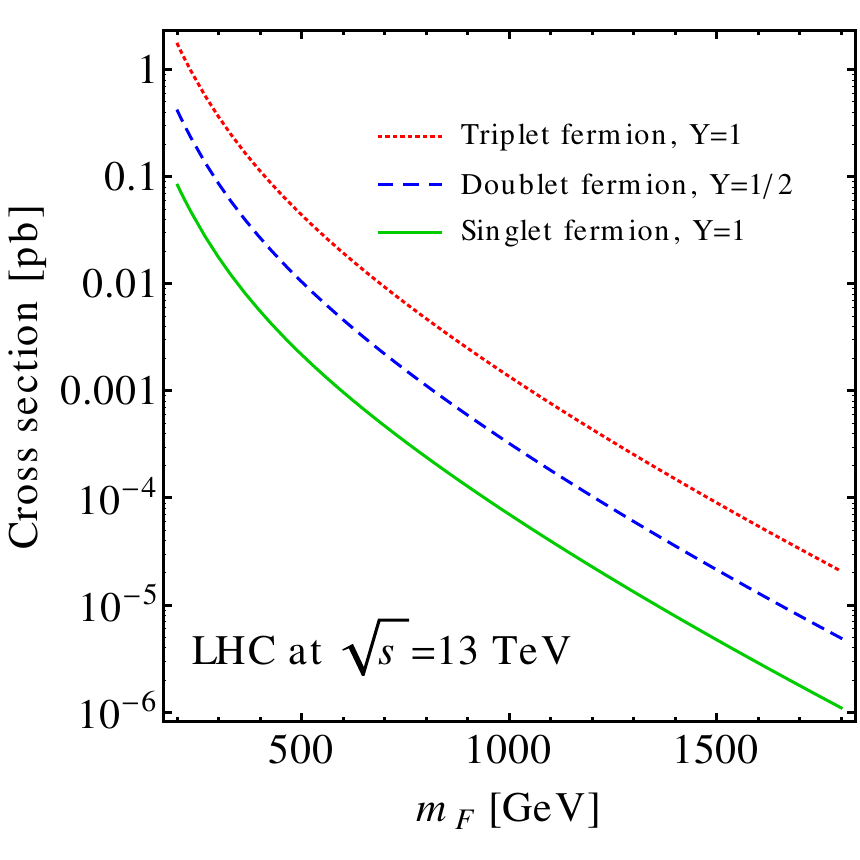}
\hspace{0.05\textwidth}
\includegraphics[width=0.40\linewidth]{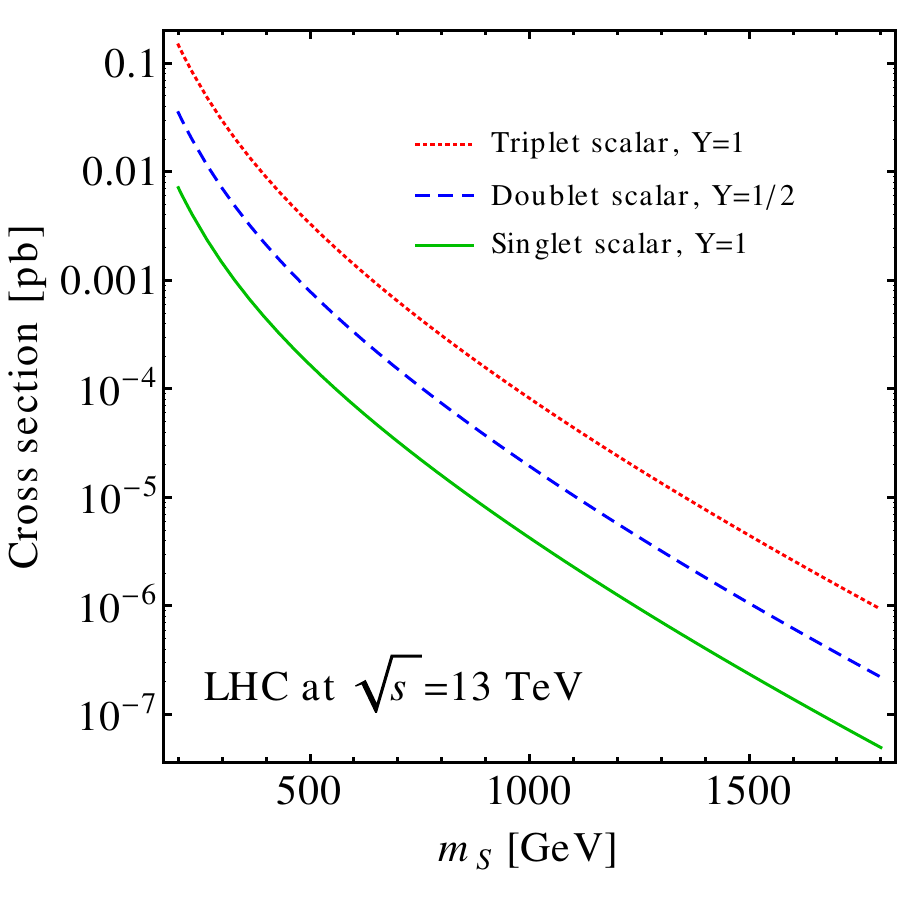}
\caption{Total hadronic cross section in $pp$ collisions at $\sqrt{s}=13$ TeV for the DY pair production of fermions (left panel) and scalars (right panel) in different $SU(2)_L$ representations: triplet (dotted red), doublet (dashed blue) and singlet (solid green). The corresponding hypercharges are indicated in the legend.}
\label{fig:cross_sec_DYF}
\end{figure}

In the left panel of \reffig{fig:cross_sec_DYF} we present the hadronic cross section at the LHC with $\sqrt{s}=13$ TeV  for the DY pair production of EW-charged colorless fermions as a function of their mass. Solid green line indicates an $SU(2)_L$ singlet with the hypercharge $Y_F=1$, dashed blue line an $SU(2)_L$ doublet with $Y_F=\frac{1}{2}$, and red dotted line an $SU(2)_L$ triplet with $Y_F=1$.  

\paragraph{Scalars}

In the case of spin-zero particles, the relevant Lagrangian for the EW interactions reads
\be
\mathcal{L}_{\text{int}} = i e\,Q_S A_\mu \left[S^{*}\partial^{\mu}S-(\partial^{\mu}S^{*})S\right]+i\frac{g_2}{\cos\theta_W}c^{\,S}Z_\mu \left[S^{*}\partial^{\mu}S_{}-(\partial^{\mu}S^{*})S\right],
\ee
where $Q_S$ denotes the scalar electric charge, $Q_S=T_{3S}+Y_S$.
Analogously to the fermion case, the coupling between an EW-charged scalar and the Z gauge boson is defined as
\bea
c^S=\left(T_{3S}\cos^2\theta_W - Y_S\sin^2\theta_{W}\right).
\eea
Likewise, three contributions to the averaged squared amplitude of the process $q\,\bar{q}\to S\,S^\ast$ are given by
\bea
		\overline{|\mathcal{M}_{\gamma}|^{2}}
		&=&  \frac{8(\pi\alpha Q_{q}Q_{S})^2}{3}\beta_{S}^2\,\sin^{2}\theta,\\
		2\,\text{Re}\overline{\left( \mathcal{M}_{\gamma}\mathcal{M}_{Z}^{\dagger}\right)}&=&  \frac{8\hat{s}\pi^2 \alpha^2 Q_{q}Q_{S} }{3\cos^2\theta_{W}\sin^2\theta_{W}}\times \frac{(\hat{s}-M_{Z}^{2})}{(\hat{s}-M_{Z}^{2})^{2}+(\Gamma M_{Z})^{2}}\,\beta_{S}^2\,\sin^{2}\theta\,c^S(c_{L}^{q} + c_{R}^{q}),\\
			\overline{|\mathcal{M}_{Z}|^{2}}  	
		&=& \frac{4(\hat{s}\pi \alpha)^2 (c^S)^2 (|c_{L}^{q}|^{2} + |c_{R}^{q}|^{2})\,\beta_{S}^2\,\sin^{2}\theta}{3\cos^{4}\theta_{W}\sin^{4}\theta_{W}{\left[{(\hat{s}- M_{Z}^{2})^2} + (\Gamma M_{Z})^2\right]}}.
\eea
Finally, the total partonic cross section for the DY production of the EW-charged scalar particles reads
\bea
			\hat{\sigma}_{q\,\bar{q}\to S\,S^\ast}(\hat{s}) &=&{\pi\alpha^2\beta_{S}^{3}}\Bigg\{	\frac{\,Q^{2}_{q}Q^{2}_{S}}{9\hat{s}} + \frac{ Q_{q}Q_{S}\,c^{S} }{9\cos^2\theta_{W}\sin^2\theta_{W}} \frac{(c_{L}^{q} + c_{R}^{q})(\hat{s}-M_{Z}^{2})}{(\hat{s}-M_{Z}^{2})^{2}+(\Gamma M_{Z})^{2}}\\
&+&\frac{\hat{s}(c^{S})^{2}(|c_{L}^{q}|^{2} + |c_{R}^{q}|^{2})}{18\cos^{4}\theta_{W}\sin^{4}\theta_{W}{\left[{(\hat{s}- M_{Z}^{2})^2} + (\Gamma M_{Z})^2\right]}}\Bigg\}.\nonumber
\eea

A comparison of the total hadronic cross sections for the EW pair production $q\,\bar{q}\to S\,S^\ast$ of colorless scalars in different representations of  $SU(2)_L$ is presented in the right panel of \reffig{fig:cross_sec_DYF}.

\section{List of the PGU models and experimental constraints}\label{app:list}

\setlength{\tabcolsep}{3pt}

\begin{scriptsize}
\begin{center}
\begin{longtable}{|c|c||c|c|c||c|c|c|c||c|}
\caption{Scenarios with one VL fermion representation and one scalar representation that allow for the  PGU ($\epsilon_{\textrm{GUT}}\leq 1\%$) and the associated unification scale lies in the range $10^{15}-10^{18}\gev$. The VL masses vary between 0.25\tev\ and 10\tev. In column 2 type of the model spectrum is given, as defined in \refeq{mass_hier}. In columns 3, 4 and 5 current lower bounds on the BSM  masses (in TeV) are presented, as provided by the measurement of the running coupling $g_3$\cite{ATLAS-CONF-2020-025}, R-hadrons\cite{Aaboud:2019trc} and charged HCSP\cite{Aaboud:2019trc} searches, respectively. In column 6 the maximal proton lifetime (in years) is quoted in the PGU region of a given model. Projected sensitivities to the EWPO\cite{Farina:2016rws} tests, the measurement of the running coupling $g_2$\cite{Alves:2014cda} and determination of $g_3$\cite{Mangano:2016jyj} at the future 100\tev\ collider are collected in columns 7, 8 and 9. Finally, in column 10 we indicate whether a given model is excluded by the current experimental data, or if it will be entirely tested in the future. The experimental bound responsible for the exclusion (testability) of a model is indicated with an asterisk.   
 }
\label{2reps_FS}\\

\hline
\multicolumn{2}{|c||}{} & \multicolumn{3}{c||}{Present limits} & \multicolumn{4}{c||}{Projected limits} & \\
\hline FS model & Type  & running $g_3$ & R-hadrons & HSCP & $\tau_p^{max}$ & EWPO & running $g_2$ & running $g_3$ & Status \\ \hline 
\endfirsthead

\multicolumn{10}{c}%
{{\bfseries \tablename\ \thetable{} -- continued from previous page}} \\
\hline
\multicolumn{2}{|c||}{} & \multicolumn{3}{c||}{Present limits} & \multicolumn{4}{c||}{Projected limits} & \\
\hline FS model & Type  & running $g_3$ & R-hadrons & HSCP & $\tau_p^{max}$ & EWPO & running $g_2$ & running $g_3$ & Status \\ \hline 
\endhead

\hline \multicolumn{10}{|r|}{{Continued on next page}} \\ \hline
\endfoot 

\endlastfoot

\hline
FS1$^{(16,6)}$ & H2 & \;\;--\;\;\quad 1.70 & \;\;--\;\;\quad 2.11 & 1.37$^*$\quad\;--\;\; & \;\;$2.3\times 10^{34\ast}$ & 2.96$^*$\quad\;--\;\; & 3.77$^*$\quad\;--\;\; & \;\;--\;\;\quad 9.00 & excluded \\
FS1$^{(18,6)}$ & H0 & \;\;--\;\;\quad 1.70 & \;\;--\;\;\quad 2.11 & 1.39 \quad\;\;--\;\; & \;\;$1.9\times 10^{35\ast}$ & 3.13 \quad\;\;--\;\; & 4.08 \quad\;\;--\;\; & \;\;--\;\;\quad 9.00 & testable \\
FS1$^{(20,6)}$ & H1 & \;\;\;--\;\;\quad 1.70$^*$ & \;\;--\;\;\quad 2.11 & 1.41 \quad\;\;--\;\; & $2.6\times 10^{36}$ & 3.30 \quad\;\;--\;\; & 4.25 \quad\;\;--\;\; & \;\;--\;\;\quad 9.00 & excluded \\
FS1$^{(22,7)}$ & H1 & \;\;--\;\;\quad 1.80 & \;\;--\;\;\quad 2.26 & 1.42 \quad\;\;--\;\; & $1.0\times 10^{38}$ & 3.46 \quad\;\;--\;\; & 4.34 \quad\;\;--\;\; & \;\;\;\;--\;\;\quad 9.00$^*$  & testable \\
FS2$^{(4,2)}$\; & H2 & \;\;--\;\;\quad 1.40 & \;\;--\;\;\quad 1.94 & 1.12$^\ast$\quad\;--\;\; & $4.9\times 10^{37}$ & 1.49$^*$ \; 2.10 & 1.19$^*$ \; 2.51 &  \;\;--\;\;\quad 8.50 & excluded \\
FS3$^{(2,13)}$ & H1 & \;\;--\;\;\quad 0.60 & \;\;\;--\;\;\quad 1.76$^\ast$ & 1.22 \quad\;\;--\;\; & $3.3\times 10^{35}$ & 2.10 \quad\;\;--\;\; & 2.51 \quad\;\;--\;\; & \;\;\;\;--\;\;\quad 7.00$^*$ & excluded \\
FS3$^{(2,14)}$ & H3 & \;\;--\;\;\quad 0.70 & \;\;--\;\;\quad 1.78 & 1.22 \quad\;\;--\;\; & $1.3\times 10^{36}$ & 2.10 \quad\;\;--\;\; & 2.51 \quad\;\;--\;\; & \;\;--\;\;\quad 7.50 & -- \\
FS3$^{(2,15)}$ & H3 & \;\;--\;\;\quad 0.80 & \;\;--\;\;\quad 1.80 & 1.22 \quad\;\;--\;\; & $4.2\times 10^{36}$ & 2.10 \quad\;\;--\;\; & 2.51 \quad\;\;--\;\; & \;\;--\;\;\quad 7.50 & -- \\
FS3$^{(2,16)}$ & H2 & \;\;--\;\;\quad 0.80 & \;\;--\;\;\quad 1.82 & 1.22 \quad\;\;--\;\; & $7.5\times 10^{36}$ & 2.10 \quad\;\;--\;\; & 2.51 \quad\;\;--\;\; & \;\;--\;\;\quad 7.50 & -- \\
FS3$^{(2,17)}$ & H2 & \;\;--\;\;\quad 0.90 & \;\;--\;\;\quad 1.84 & 1.22$^\ast$\quad\;--\;\; & $7.5\times 10^{36}$ & 2.10$^*$\quad\;--\;\; & 2.51$^*$\quad\;--\;\; & \;\;--\;\;\quad 7.50 & excluded \\
FS3$^{(3,22)}$ & H1 & \;\;\;--\;\;\quad 1.20$^*$ & \;\;--\;\;\quad 1.89 & 1.30 \quad\;\;--\;\; & $3.5\times 10^{41}$ & 2.56 \quad\;\;--\;\; & 3.23 \quad\;\;--\;\; & \;\;--\;\;\quad 8.00 & excluded \\
FS3$^{(3,23)}$ & H1 & \;\;--\;\;\quad 1.20 & \;\;\;--\;\;\quad 1.90$^\ast$  & 1.30 \quad\;\;--\;\; & $2.0\times 10^{42}$ & 2.56 \quad\;\;--\;\; & 3.23 \quad\;\;--\;\; & \;\;\;\;--\;\;\quad 8.00$^*$ & excluded \\
FS3$^{(3,24)}$ & H1 & \;\;--\;\;\quad 1.30 & \;\;--\;\;\quad 1.91 & 1.30 \quad\;\;--\;\; & $1.1\times 10^{43}$ & 2.56 \quad\;\;--\;\; & 3.23 \quad\;\;--\;\; & \;\;\;\;--\;\;\quad 8.50$^*$  & testable \\
FS3$^{(3,25)}$ & H0 & \;\;--\;\;\quad 1.30 & \;\;--\;\;\quad 1.91 & 1.30 \quad\;\;--\;\; & $5.4\times 10^{43}$ & 2.56 \quad\;\;--\;\; & 3.23 \quad\;\;--\;\; & \;\;--\;\;\quad 8.50 & -- \\
FS3$^{(3,26)}$ & H2 & \;\;--\;\;\quad 1.40 & \;\;--\;\;\quad 1.92 & 1.30 \quad\;\;--\;\; & $1.3\times 10^{44}$ & 2.56 \quad\;\;--\;\; & 3.23 \quad\;\;--\;\; & \;\;--\;\;\quad 8.50 & -- \\
FS3$^{(3,27)}$ & H2 & \;\;--\;\;\quad 1.40 & \;\;--\;\;\quad 1.93 & 1.30 \quad\;\;--\;\; & $1.5\times 10^{44}$ & 2.56$^\ast$ \quad\;--\;\; & 3.23$^\ast$ \quad\;--\;\; & \;\;--\;\;\quad 8.50 & testable \\
FS4$^{(2,3)}$\; & H3 & \;\;--\;\;\quad 0.90 & \;\;--\;\;\quad 1.87 & 1.22 \quad\;\;--\;\; & $1.1\times 10^{42}$ & 2.10 \quad\;\;--\;\; & 2.51 \quad\;\;--\;\; & \;\;--\;\;\quad 7.50 & -- \\
FS5$^{(3,4)}$\; & H0 & \;\;--\;\;\quad 1.20 & \;\;--\;\;\quad 1.94 & 1.30 \quad\;\;--\;\; & $3.4\times 10^{35}$ & 2.56 \quad\;\;--\;\; & 3.23 \quad\;\;--\;\; & \;\;--\;\;\quad 8.00 & -- \\
FS5$^{(4,5)}$\; & H1 & \;\;\;--\;\;\quad 1.50$^*$ & \;\;--\;\;\quad 1.98 & 1.34 \quad\;\;--\;\; & $6.6\times 10^{37}$ & 2.96 \quad\;\;--\;\; & 3.77 \quad\;\;--\;\; & \;\;--\;\;\quad 9.00 & excluded \\
FS5$^{(4,6)}$\; & H2 & \;\;--\;\;\quad 1.70 & \;\;--\;\;\quad 2.11 & 1.34 \quad\;\;--\;\; & $1.1\times 10^{39}$ & 2.06$^\ast$ \quad\;--\;\; & 3.77$^\ast$ \quad\;--\;\; & \;\;--\;\;\quad 9.00 & testable \\
FS5$^{(5,7)}$\; & H0 & \;\;--\;\;\quad 1.80 & \;\;--\;\;\quad 2.26 & 1.39 \quad\;\;--\;\; & $4.0\times 10^{42}$ & 3.30 \quad\;\;--\;\; & 4.25 \quad\;\;--\;\; & \;\;--\;\;\quad 9.00 & -- \\
\;FS6$^{(10,2)}$ & H2 & 1.00 \quad\;\;--\;\; & 1.87 \quad\;\;--\;\; & \;\;--\;\;\quad 0.97 & $2.8\times 10^{37}$ & \;\;--\;\;\quad 2.34 & \;\;--\;\;\quad 2.92 & 8.00$^*$\quad\;--\;\; & testable \\
\;FS6$^{(12,2)}$ & H1 & 1.20 \quad\;\;--\;\; & 1.93 \quad\;\;--\;\; & \;\;\;--\;\;\quad 0.97$^\ast$ & $3.5\times 10^{38}$ & \;\;\;\;--\;\;\quad 2.34$^*$ & \;\;\;\;--\;\;\quad 2.92$^*$ & 8.50 \quad\;\;--\;\; & excluded \\
FS7$^{(2,4)}$\; & H3 & 0.25 \quad 0.25 & 1.56 \quad 1.57 & \;\;--\;\;\quad\;\;--\;\;\ & $3.1\times 10^{37}$ & \;\;--\;\;\quad 1.82 & \;\;--\;\;\quad 1.97 & 3.00 \quad 6.00 & -- \\
FS7$^{(4,6)}$\; & H3 & 0.25 \quad 0.60 & 1.69 \quad 1.73 & \;\;--\;\;\quad\;\;--\;\;\ & $1.7\times 10^{44}$ & \;\;--\;\;\quad 2.22 & \;\;--\;\;\quad 2.73 & 5.50 \quad 7.00 & -- \\ 
FS8$^{(4,1)}$\; & H3 & 0.25 \quad 0.25 & 1.69 \quad 1.38 & \;\;--\;\;\quad\;\;--\;\;\ & \;\;$2.3\times 10^{34\ast}$ & \;\;--\;\;\quad 1.82 & \;\;--\;\;\quad 1.97 & 5.50 \quad 2.50 & testable \\
FS9$^{(10,2)}$ & H1 & 1.00 \quad 0.25 & 1.87 \quad 1.49 & \;\;--\;\;\quad\;\;--\;\;\ & \;\;$1.5\times 10^{35\ast}$ & \;\;\;--\;\;\quad 2.56$^\ast$ & \;\;\;--\;\;\quad 3.23$^\ast$ & \;\;8.00 \quad 4.50$^*$ & testable \\
FS9$^{(16,3)}$ & H0 & 1.60 \quad 0.80 & 2.01 \quad 1.62 & \;\;--\;\;\quad\;\;--\;\;\ & $4.3\times 10^{42}$ & \;\;--\;\;\quad 3.13 & \;\;--\;\;\quad 4.15 & 8.50 \quad 5.00 & -- \\
FS10$^{(2,5)}$ & H3 & 0.25 \quad 0.40 & 1.56 \quad  1.67 & \;\;--\;\;\quad\;\;--\;\;\ & \;\;$7.2\times 10^{34\ast}$ & \;\;--\;\;\quad 2.03 & \;\;--\;\;\quad 2.39 & 3.50 \quad 6.00 & testable \\
FS10$^{(4,7)}$ & H1 & 0.25 \quad 0.70 & \;\;1.69 \quad 1.78$^*$ & \;\;--\;\;\quad\;\;--\;\;\ & $4.4\times 10^{36}$ & \;\;\;\;--\;\;\quad 2.40$^*$ & \;\;\;\;--\;\;\quad 3.00$^*$ & \;\;4.00 \quad 7.00$^*$ & excluded\\
FS10$^{(4,8)}$ & H3 & 0.25 \quad 0.80 & 1.69 \quad 1.82 & \;\;--\;\;\quad\;\;--\;\;\ & $3.9\times 10^{37}$ & \;\;--\;\;\quad 2.56 & \;\;--\;\;\quad 3.23 & 4.00 \quad 7.00 & -- \\
\;\,FS10$^{(6,10)}$ & H1 & 0.50 \quad 1.10 & 1.76 \quad 1.87 & \;\;--\;\;\quad\;\;--\;\;\ & $4.6\times 10^{40}$ & \;\;--\;\;\quad 2.86 & \;\;\;--\;\;\quad 3.66$^\ast$ & \;\;6.50 \quad 8.50$^*$ &  testable\\
\;\,FS10$^{(6,11)}$ & H2 & 0.50 \quad 1.10 & 1.76$^\ast$ \; 1.89 & \;\;--\;\;\quad\;\;--\;\;\ & $4.5\times 10^{40}$ & \;\;--\;\;\quad 3.00 & \;\;--\;\;\quad 3.82 & 6.50$^*$ \; 8.50 & excluded \\
FS11$^{(6,1)}$ & H1 & \;\;0.50 \quad 0.90$^*$ & 1.76 \quad 1.87 & \;\;--\;\;\quad\;\;--\;\;\ & $8.2\times 10^{36}$ & \;\;--\;\;\quad 2.56 & \;\;--\;\;\quad 3.23 & 6.50 \quad 7.50 & excluded \\
FS12$^{(2,2)}$ & H1 & 0.25 \quad 0.25 & \;\;1.69 \quad 1.31$^*$  & \;\;--\;\;\quad\;\;--\;\; & \;$8.9\times 10^{34\ast}$ & 1.82 \quad\;\;--\;\; & 1.97 \quad\;\;--\;\; & \;\;5.50 \quad 4.00$^*$ & testable\\
FS12$^{(2,3)}$ & H3 & 0.25 \quad 0.25 & 1.69 \quad 1.38 & \;\;--\;\;\quad\;\;--\;\; & $4.1\times 10^{36}$ & 1.82 \quad\;\;--\;\; & 1.97 \quad\;\;--\;\; & 5.50 \quad 4.00 & -- \\
FS12$^{(2,4)}$ & H2 & 0.25 \quad 0.25 & 1.69 \quad 1.44 & \;\;--\;\;\quad\;\;--\;\; & $1.2\times 10^{37}$ & 1.82 \quad\;\;--\;\; & 1.97$^\ast$\quad\;--\;\; & 5.50$^*$ \; 4.00 & testable \\
FS13$^{(4,7)}$ & H3 & 0.80 \quad 0.25 & 1.80 \quad 1.51 & \;\;--\;\;\quad\;\;--\;\; & $1.9\times 10^{37}$ & 2.56 \quad\;\;--\;\; & 3.23 \quad\;\;--\;\; & 7.50 \quad 5.50 & -- \\
FS13$^{(4,8)}$ & H2 & 0.80 \quad 0.25 & 1.93$^\ast$ \; 1.57 & \;\;--\;\;\quad\;\;--\;\; & $4.8\times 10^{37}$ & 2.56$^*$ \quad\;--\;\; & 3.23$^*$ \quad\;--\;\; & 7.50$^*$ \; 5.50 & excluded \\
\;\;FS13$^{(6,11)}$ & H1 & \;\;1.20 \quad 0.50$^*$ & 1.93 \quad 1.70 & \;\;--\;\;\quad\;\;--\;\; & $6.4\times 10^{40}$ & 3.13 \quad\;\;--\;\; & 4.15 \quad\;\;--\;\; & 8.50 \quad 6.50 & excluded \\
\;\;FS13$^{(6,12)}$ & H1 & 1.20 \quad 0.50 & 1.93 \quad 1.73 & \;\;--\;\;\quad\;\;--\;\; & $1.5 \times 10^{42}$ & 3.13 \quad\;\;--\;\; & 4.15 \quad\;\;--\;\; & \;\;8.50 \quad 6.50$^*$ & testable \\
\;\;FS13$^{(6,13)}$ & H2 & 1.30 \quad 0.70 & 1.93 \quad 1.76 & \;\;--\;\;\quad\;\;--\;\; & $5.7\times 10^{42}$ & 3.13 \quad\;\;--\;\; & 4.15$^\ast$ \quad\;--\;\; & 8.50$^*$ \; 7.00 & testable \\
FS14$^{(4,2)}$ & H2 & 1.10 \quad 1.30 & 1.80 \quad 1.94 & \;\;--\;\;\quad\;\;--\;\; & $1.3\times 10^{43}$ & 2.56 \quad 1.82 & 3.23 \quad 1.97 & 7.50$^*$ \; 8.50 & testable\\
FS15$^{(4,1)}$ & H1 & 0.80 \quad 0.25 & 1.80 \quad 1.63 & \;\;--\;\;\quad\;\;--\;\; & $9.3\times 10^{36}$ & 2.56 \quad\;\;--\;\; & 3.23 \quad\;\;--\;\; & \;\;7.50 \quad 5.00$^*$ & testable \\
FS15$^{(6,2)}$ & H2 & 1.30$^\ast$ \; 0.70 & 1.93 \quad 1.76 & \;\;--\;\;\quad\;\;--\;\; & $4.1\times 10^{42}$ & 3.13 \quad\;\;--\;\; & 4.15 \quad\;\;--\;\; & 8.50 \quad 7.00 & excluded \\
\;\;FS16$^{(4,39)}$ & H1 & 1.90 \quad\;\;--\;\; & 2.16 \quad\;\;--\;\; & \;\;\;\;--\;\;\quad 1.11$^\ast$ & $6.8\times 10^{36}$ & \;\;\;\;--\;\;\quad 3.26$^*$ & \;\;\;\;--\;\;\quad 4.23$^*$ & 9.00 \quad\;\;--\;\; & excluded \\
\;\;FS16$^{(4,40)}$ & H1 & 1.90 \quad\;\;--\;\; & 2.16 \quad\;\;--\;\; & \;\;--\;\;\quad 1.12 & $2.0\times 10^{37}$ & \;\;\;\;--\;\;\quad 3.30$^\ast$ & \;\;\;\;--\;\;\quad 4.25$^\ast$ & 9.00 \quad\;\;--\;\; & testable \\ 
\;\;FS16$^{(4,41)}$ & H1 & 1.90 \quad\;\;--\;\; & 2.16 \quad\;\;--\;\; & \;\;--\;\;\quad 1.12 & $6.2\times 10^{37}$ & \;\;\;\;--\;\;\quad 3.34$^\ast$ & \;\;\;\;--\;\;\quad 4.27$^\ast$ & 9.00 \quad\;\;--\;\; & testable \\
\;\;FS16$^{(4,42)}$ & H0 & 1.90 \quad\;\;--\;\; & 2.16 \quad\;\;--\;\; & \;\;--\;\;\quad 1.12 & $5.9\times 10^{37}$ & \;\;--\;\;\quad 3.38 & \;\;--\;\;\quad 4.30 & 9.00 \quad\;\;--\;\; & -- \\
\;\;FS16$^{(4,43)}$ & H0 & 1.90 \quad\;\;--\;\; & 2.16 \quad\;\;--\;\; & \;\;--\;\;\quad 1.12 & $3.5\times 10^{37}$ & \;\;--\;\;\quad 3.42 & \;\;--\;\;\quad 4.32 & 9.00 \quad\;\;--\;\; & -- \\
\;\;FS16$^{(4,44)}$ & H2 & 1.90 \quad\;\;--\;\; & 2.16 \quad\;\;--\;\; & \;\;--\;\;\quad 1.13 & $8.2\times 10^{37}$ & \;\;--\;\;\quad 3.46 & \;\;--\;\;\quad 4.34 & 9.00$^*$\quad\;\;--\;\; & testable \\
\;\;FS16$^{(4,45)}$ & H2 & 1.90 \quad\;\;--\;\; & 2.16 \quad\;\;--\;\; & \;\;--\;\;\quad 1.13 & $1.0\times 10^{39}$ & \;\;--\;\;\quad 3.50 & \;\;--\;\;\quad 4.37 & 9.00$^*$\quad\;\;--\;\; & testable \\
\;\;FS16$^{(4,46)}$ & H2 & 1.90 \quad\;\;--\;\; & 2.16$^\ast$ \quad\;--\;\; & \;\;--\;\;\quad 1.13 & $9.6\times 10^{38}$ & \;\;--\;\;\quad 3.54 & \;\;--\;\;\quad 4.41 & 9.00$^*$ \quad\;--\;\; & excluded \\
\;\;FS16$^{(4,47)}$ & H2 & 1.90$^\ast$ \quad\;--\;\; & 2.16 \quad\;\;--\;\; & \;\;--\;\;\quad 1.13 & $9.9\times 10^{38}$ & \;\;--\;\;\quad 3.58 & \;\;--\;\;\quad 4.45 & 9.00 \quad\;\;--\;\; & excluded \\
FS17$^{(2,2)}$ & H0 & 1.10 \quad 0.25 & 2.07 \quad 1.49 & \;\;--\;\;\quad\;\;--\;\; & $5.6\times 10^{41}$ & \;\;--\;\;\quad 2.56 & \;\;--\;\;\quad 3.23 & 8.50 \quad 5.00 & -- \\
FS18$^{(4,4)}$ & H2 & 1.90$^\ast$ \quad\;--\;\; & 2.16 \quad\;\;--\;\; & \;\;--\;\;\quad 1.09 & $2.4\times 10^{40}$ & \;\;--\;\;\quad 3.30  & \;\;--\;\;\quad 4.25 & 9.00 \quad\;\;--\;\; & excluded \\
FS19$^{(2,2)}$ & H1 & 1.10 \quad 0.25 & 2.07 \quad 1.49 & \;\;--\;\;\quad\;\;--\;\; & $4.0\times 10^{39}$ & \;\;--\;\;\quad 2.56  & \;\;--\;\;\quad 3.23 & \;\;8.50 \quad 5.00$^*$ & testable \\
\;\;FS20$^{(2,23)}$ & H1 & 1.30 \quad\;\;--\;\; & 2.16 \quad\;\;--\;\; & \;\;\;--\;\;\quad 1.03$^\ast$ & $1.6\times 10^{36}$ & \;\;\;\;--\;\;\quad 2.51$^*$ & \;\;\;\;--\;\;\quad 3.16$^*$ & 8.00 \quad\;\;--\;\; & excluded \\
\;\;FS20$^{(2,24)}$ & H1 & 1.30 \quad\;\;--\;\; & 2.16 \quad\;\;--\;\; & \;\;--\;\;\quad 1.04 & $7.0\times 10^{36}$ & \;\;\;\;--\;\;\quad 2.56$^\ast$ & \;\;\;\;--\;\;\quad 3.23$^\ast$ & 8.00 \quad\;\;--\;\; & testable \\
\;\;FS20$^{(2,25)}$ & H0 & 1.30 \quad\;\;--\;\; & 2.16 \quad\;\;--\;\; & \;\;--\;\;\quad 1.05 & $3.4\times 10^{37}$ & \;\;--\;\;\quad 2.62 & \;\;--\;\;\quad 3.31 & 8.00 \quad\;\;--\;\; & -- \\
\;\;FS20$^{(2,26)}$ & H0 & 1.30 \quad\;\;--\;\; & 2.16 \quad\;\;--\;\; & \;\;--\;\;\quad 1.05 & $2.8\times 10^{37}$ & \;\;--\;\;\quad 2.67 & \;\;--\;\;\quad 3.39 & 8.00 \quad\;\;--\;\; & -- \\
\;\;FS20$^{(2,27)}$ & H2 & 1.30 \quad\;\;--\;\; & 2.16 \quad\;\;--\;\; & \;\;--\;\;\quad 1.06 & $3.3\times 10^{37}$ & \;\;--\;\;\quad 2.72 & \;\;--\;\;\quad 4.47 & 8.00$^*$\quad\;--\;\; & testable \\
\;\;FS20$^{(2,28)}$ & H2 & 1.30$^\ast$ \quad\;--\;\; & 2.16 \quad\;\;--\;\; & \;\;--\;\;\quad 1.07 & $3.5\times 10^{37}$ & \;\;--\;\;\quad 2.77 & \;\;--\;\;\quad 3.54 & 8.00 \quad\;\;--\;\; & excluded \\
\;\;FS20$^{(3,35)}$ & H0 & 1.80 \quad\;\;--\;\; & 2.19 \quad\;\;--\;\; & \;\;--\;\;\quad 1.10 & $1.3\times 10^{45}$ & \;\;--\;\;\quad 3.09 & \;\;--\;\;\quad 4.10 & 9.00 \quad\;\;--\;\; & -- \\
\;\;FS20$^{(3,36)}$ & H0 & 1.80 \quad\;\;--\;\; & 2.19 \quad \;\;--\;\; & \;\;--\;\;\quad 1.10 & $2.7\times 10^{45}$ & \;\;--\;\;\quad 3.13 & \;\;--\;\;\quad 4.15 & 9.00 \quad\;\;--\;\; & -- \\
FS21$^{(1,1)}$ & H1 & 0.60 \quad\;\;--\;\; & 2.07 \quad\;\;--\;\; & \;\;\;--\;\;\quad 0.88$^\ast$ & $8.6\times 10^{36}$ & \;\;\;\;--\;\;\quad 1.66$^*$ & \;\;\;\;--\;\;\quad 1.60$^*$ & 7.00 \quad\;\;--\;\; & excluded \\
FS22$^{(1,1)}$ & H1 & 0.60 \quad 0.25 & 2.07 \; 1.38$^\ast$ & \;\;--\;\;\quad\;\;--\;\; & $1.3\times 10^{41}$ & \;\;--\;\;\quad 1.82  & \;\;--\;\;\quad 1.97 & \;\;7.00 \quad 3.00$^*$ & excluded \\
\hline
\end{longtable}
\end{center}
\end{scriptsize}

\begin{scriptsize}
\begin{center}
\begin{longtable}{|c|c||c|c|c||c|c|c|c||c|}
\caption{Scenarios with two different representations of scalars that allow for the  PGU ($\epsilon_{\textrm{GUT}}\leq 1\%$) and the associated unification scale lies in the range $10^{15}-10^{18}\gev$. The scalar masses vary between 0.25\tev\ and 10\tev. In column 2 type of the model spectrum is given, as defined in \refeq{mass_hier}. In columns 3, 4 and 5 current lower bounds on the BSM  masses (in TeV) are presented, as provided by the measurement of the running coupling $g_3$\cite{ATLAS-CONF-2020-025}, R-hadrons\cite{Aaboud:2019trc} and charged HCSP\cite{Aaboud:2019trc} searches, respectively. In column 6 the maximal proton lifetime (in years) is quoted in the PGU region of a given model. Projected sensitivities to the EWPO\cite{Farina:2016rws} tests, the measurement of the running coupling $g_2$\cite{Alves:2014cda} and determination of $g_3$\cite{Mangano:2016jyj} at the future 100\tev\ collider are collected in columns 7, 8 and 9. Finally, in column 10 we indicate whether a given model is excluded by the current experimental limits, or if it will be entirely tested in the future. The experimental bound responsible for the exclusion (testability)of a model is indicated with an asterisk.}
\label{2reps_SS}\\

\hline
\multicolumn{2}{|c||}{} & \multicolumn{3}{c||}{Present limits} & \multicolumn{4}{c||}{Projected limits} & \\
\hline SS model & Type  & running $g_3$ & R-hadrons & HSCP &  $\tau_p^{max}$ & EWPO & running $g_2$ & running $g_3$ & Status \\ \hline 
\endfirsthead

\multicolumn{10}{c}%
{{\bfseries \tablename\ \thetable{} -- continued from previous page}} \\
\hline
\multicolumn{2}{|c||}{} & \multicolumn{3}{c||}{Present limits} & \multicolumn{4}{c||}{Projected limits} & \\
\hline SS model & Type & running $g_3$ & R-hadrons & HSCP &  $\tau_p^{max}$ & EWPO & running $g_2$ & running $g_3$ & Status \\ \hline 
\endhead

\hline \multicolumn{10}{|r|}{{Continued on next page}} \\ \hline
\endfoot 

\endlastfoot

\hline
SS1$^{(40,6)}$ & H1 & \;\;\;\;--\;\;\quad 1.70$^*$ & \;\;--\;\;\quad 2.03 & 1.11 \quad\;\;--\;\; & $1.8\times 10^{35}$ & 3.30 \quad\;\;--\;\; & 4.25 \quad\;\;--\;\; & \;\;--\;\;\quad 8.50 & excluded \\
SS1$^{(41,6)}$ & H1 & \;\;\;\;--\;\;\quad 1.70$^*$ & \;\;--\;\;\quad 2.03 & 1.12 \quad\;\;--\;\; & $5.0\times 10^{35}$ & 3.34 \quad\;\;--\;\; & 4.27 \quad\;\;--\;\; & \;\;--\;\;\quad 8.50 & excluded \\
SS1$^{(42,6)}$ & H1 & \;\;\;\;--\;\;\quad 1.70$^*$ & \;\;--\;\;\quad 2.03 & 1.12 \quad\;\;--\;\; & $5.3\times 10^{35}$ & 3.38 \quad\;\;--\;\; & 4.29 \quad\;\;--\;\; & \;\;--\;\;\quad 8.50 & excluded \\
SS1$^{(43,7)}$ & H0 & \;\;--\;\;\quad 1.80 & \;\;--\;\;\quad 2.09 & 1.12 \quad\;\;--\;\; & $9.4\times 10^{36}$ & 3.42 \quad\;\;--\;\; & 4.32 \quad\;\;--\;\; & \;\;\;\;--\;\;\quad 9.00$^*$ & testable \\
SS1$^{(44,7)}$ & H1 & \;\;--\;\;\quad 1.80 & \;\;--\;\;\quad 2.09 & 1.13 \quad\;\;--\;\; & $9.8\times 10^{36}$ & 3.46 \quad\;\;--\;\; & 4.34 \quad\;\;--\;\; & \;\;\;\;--\;\;\quad 9.00$^*$ & testable \\
SS1$^{(45,7)}$ & H1 & \;\;--\;\;\quad 1.80 & \;\;--\;\;\quad 2.09 & 1.13 \quad\;\;--\;\; & $1.6\times 10^{37}$ & 3.50 \quad\;\;--\;\; & 4.37 \quad\;\;--\;\; & \;\;\;\;--\;\;\quad 9.00$^*$ & testable \\
SS1$^{(46,7)}$ & H1 & \;\;--\;\;\quad 1.80 & \;\;\;\;--\;\;\quad 2.09$^*$ & 1.13 \quad\;\;--\;\; & $3.3\times 10^{37}$ & 3.54 \quad\;\;--\;\; & 4.41 \quad\;\;--\;\; & \;\;\;\;--\;\;\quad 9.00$^*$ & excluded \\
SS1$^{(47,7)}$ & H1 & \;\;\;\;--\;\;\quad 1.80$^*$ & \;\;--\;\;\quad 2.09 & 1.13 \quad\;\;--\;\; & $3.2\times 10^{37}$ & 3.58 \quad\;\;--\;\; & 4.45 \quad\;\;--\;\; & \;\;--\;\;\quad 9.00 & excluded \\
SS2$^{(8,2)}$\;\; & H2 & \;\;--\;\;\quad 1.40 & \;\;--\;\;\quad 1.94 & 0.87$^*$\quad\;--\;\; & $3.4\times 10^{37}$ & 1.49$^*$ \; 2.10 & 1.19$^*$ \; 2.51 & \;\;--\;\;\quad 8.00 & excluded \\
SS2$^{(9,2)}$\;\; & H3 & \;\;--\;\;\quad 1.40 & \;\;--\;\;\quad 1.94 & 0.88 \quad\;\;--\;\; & $2.0\times 10^{39}$ & 1.58 \quad 2.10 & 1.41 \quad 2.51 & \;\;--\;\;\quad 8.00 & -- \\
SS2$^{(10,2)}$\; & H1 & \;\;\;\;--\;\;\quad 1.40$^*$ & \;\;--\;\;\quad 1.94 & 0.89 \quad\;\;--\;\; & $8.4\times 10^{38}$ & 1.66 \quad 2.10 & 1.60 \quad 2.51 & \;\;--\;\;\quad 8.00 &  excluded \\
SS3$^{(2,18)}$\; & H1 & \;\;\;\;--\;\;\quad 1.00$^*$ & \;\;--\;\;\quad 1.85 & 0.97 \quad\;\;--\;\; & $1.9\times 10^{36}$ & 2.34 \quad\;\;--\;\; & 2.92 \quad\;\;--\;\; & \;\;--\;\;\quad 7.50 & excluded \\
SS3$^{(2,19)}$\; & H1 & \;\;--\;\;\quad 1.00 & \;\;--\;\;\quad 1.86 & 0.97 \quad\;\;--\;\; & $6.5\times 10^{36}$ & 2.34 \quad\;\;--\;\; & 2.92 \quad\;\;--\;\; & \;\;\;\;--\;\;\quad 7.50$^*$ & testable \\
SS3$^{(2,20)}$\; & H0 & \;\;--\;\;\quad 1.10 & \;\;--\;\;\quad 1.87 & 0.97 \quad\;\;--\;\; & $1.5\times 10^{37}$ & 2.34 \quad\;\;--\;\; & 2.92 \quad\;\;--\;\; & \;\;--\;\;\quad 8.00 & -- \\
SS3$^{(2,21)}$\; & H0 & \;\;--\;\;\quad 1.10 & \;\;--\;\;\quad 1.88 & 0.97 \quad\;\;--\;\; & $8.2\times 10^{37}$ & 2.34 \quad\;\;--\;\; & 2.92 \quad\;\;--\;\; & \;\;--\;\;\quad 8.00 & -- \\
SS3$^{(2,22)}$\; & H2 & \;\;--\;\;\quad 1.20 & \;\;--\;\;\quad 1.89 & 0.97 \quad\;\;--\;\; & $9.7\times 10^{37}$ & 2.34 \quad\;\;--\;\; & 2.92 \quad\;\;--\;\; & \;\;--\;\;\quad 8.00 & -- \\
SS3$^{(2,23)}$\; & H2 & \;\;--\;\;\quad 1.20 & \;\;--\;\;\quad 1.90 & 0.97 \quad\;\;--\;\; & $1.0\times 10^{38}$ & 2.34$^*$\quad\;--\;\; & 2.92$^*$\quad\;--\;\; & \;\;--\;\;\quad 8.00 & testable \\
SS3$^{(2,24)}$\; & H2 & \;\;--\;\;\quad 1.30 & \;\;--\;\;\quad 1.91 & 0.97$^*$\quad\;--\;\; & $1.2\times 10^{38}$ & 2.34$^*$\quad\;--\;\; & 2.92$^*$\quad\;--\;\; & \;\;--\;\;\quad 8.00 & excluded \\
SS4$^{(4,7)}$\;\; & H1 & \;\;--\;\;\quad 1.90 & \;\;--\;\;\quad 2.09 & 1.09 \quad\;\;--\;\; & $6.7\times 10^{36}$ & 3.33 \quad\;\;--\;\; & 4.25 \quad\;\;--\;\; & \;\;\;\;--\;\;\quad 9.00$^*$ & testable \\
SS5$^{(3,4)}$\;\; & H3 & 0.25 \quad 0.25 & 1.38 \quad 1.57 & \;\;--\;\;\quad\;\;--\;\; & $2.0\times 10^{36}$ & \;\;--\;\;\quad 1.82 & \;\;--\;\;\quad 1.97 & 2.50 \quad 5.00 & -- \\
SS5$^{(4,4)}$\;\; & H3 & 0.25 \quad 0.25 & 1.44 \quad 1.57 & \;\;--\;\;\quad\;\;--\;\; & $2.7\times 10^{37}$ & \;\;--\;\;\quad 1.82 & \;\;--\;\;\quad 1.97 & 3.50 \quad 6.00 & -- \\
SS5$^{(5,4)}$\;\; & H1 & 0.25 \quad 0.25 & \;\;1.47 \quad 1.57$^*$ & \;\;--\;\;\quad\;\;--\;\; & $4.3\times 10^{37}$ & \;\;\;\;--\;\;\quad 1.82$^*$ & \;\;\;\;--\;\;\quad 1.97$^*$ & \;\;4.00 \quad 6.00$^*$ & excluded \\
SS5$^{(5,5)}$\;\; & H2 & 0.25 \quad 0.40 & 1.47 \quad 1.67 & \;\;--\;\;\quad\;\;--\;\; & $1.8\times 10^{39}$ & \;\;--\;\;\quad 2.03 & \;\;--\;\;\quad 2.40 & 4.00$^*$ \; 6.50 & testable \\
SS5$^{(6,5)}$\;\; & H3 & 0.25 \quad 0.40 & 1.49 \quad 1.67 & \;\;--\;\;\quad\;\;--\;\; & $5.5\times 10^{40}$ & \;\;--\;\;\quad 2.03 & \;\;--\;\;\quad 2.40 & 5.00 \quad 6.50 & -- \\
SS5$^{(7,5)}$\;\; & H1 & 0.25 \quad 0.40 & 1.51 \quad 1.67 & \;\;--\;\;\quad\;\;--\;\; & $2.3\times 10^{41}$ & \;\;\;\;--\;\;\quad 2.03$^*$ & \;\;\;\;--\;\;\quad 2.40$^*$ & 5.50 \quad 6.50 & testable \\
SS5$^{(7,6)}$\;\; & H2 & 0.25 \quad 0.60 & 1.51$^*$\quad 1.73 & \;\;--\;\;\quad\;\;--\;\; & $1.9\times 10^{42}$ & \;\;--\;\;\quad 2.22 & \;\;--\;\;\quad 2.73 & 5.50$^*$ \; 7.00 & excluded \\
SS5$^{(8,6)}$\;\; & H3 & 0.25 \quad 0.60 & 1.57 \quad 1.73 & \;\;--\;\;\quad\;\;--\;\; & $1.9\times 10^{44}$ & \;\;--\;\;\quad 2.22 & \;\;--\;\;\quad 2.73 & 5.50 \quad 7.00 & -- \\
SS6$^{(27,3)}$ & H2 & 1.40$^*$\quad 0.40 & 1.93 \quad 1.62 & \;\;--\;\;\quad\;\;--\;\; & $6.4\times 10^{38}$ & \;\;--\;\;\quad 3.13 & \;\;--\;\;\quad 4.15 & 8.50 \quad 7.00 & excluded \\
SS6$^{(28,3)}$ & H2 & 1.40$^*$\quad 0.40 & 1.93 \quad 1.62 & \;\;--\;\;\quad\;\;--\;\; & $2.7\times 10^{39}$ & \;\;--\;\;\quad 3.13 & \;\;--\;\;\quad 4.15 & 8.50 \quad 7.00 & excluded \\
SS6$^{(29,3)}$ & H2 & 1.50 \quad 0.40 & 1.94 \quad 1.62 & \;\;--\;\;\quad\;\;--\;\; & $1.2\times 10^{40}$ & \;\;--\;\;\quad 3.13 & \;\;--\;\;\quad 4.15 & 8.50$^*$ \; 7.00 & testable \\
SS6$^{(30,3)}$ & H0 & 1.50 \quad 0.40 & 1.94 \quad 1.62 & \;\;--\;\;\quad\;\;--\;\; & $1.8\times 10^{40}$ & \;\;--\;\;\quad 3.13 & \;\;--\;\;\quad 4.15 & 8.50 \quad 7.00 & -- \\
SS6$^{(31,3)}$ & H0 & 1.50 \quad 0.40 & 1.94 \quad 1.62 & \;\;--\;\;\quad\;\;--\;\; & $4.4\times 10^{40}$ & \;\;--\;\;\quad 3.13 & \;\;--\;\;\quad 4.15 & \;\;8.50 \quad 7.00$^*$ & testable \\
SS6$^{(32,3)}$ & H1 & 1.60 \quad 0.40 & 1.95 \quad 1.62 & \;\;--\;\;\quad\;\;--\;\; & $6.8\times 10^{40}$ & \;\;--\;\;\quad 3.13 & \;\;\;\;--\;\;\quad 4.15$^*$ & \;\;8.50 \quad 7.00$^*$ & testable \\
SS6$^{(33,3)}$ & H1 & 1.60 \quad 0.40 & 1.95 \quad 1.62 & \;\;--\;\;\quad\;\;--\;\; & $9.7\times 10^{40}$ & \;\;\;\;--\;\;\quad 3.13$^*$ & \;\;\;\;--\;\;\quad 4.15$^*$ & \;\;8.50 \quad 7.00$^*$ & testable \\
SS6$^{(34,3)}$ & H1 & 1.60 \quad 0.40 & \;\;1.96 \quad 1.62$^*$ & \;\;--\;\;\quad\;\;--\;\; & $5.3\times 10^{40}$ & \;\;\;\;--\;\;\quad 3.13$^*$ & \;\;\;\;--\;\;\quad 4.15$^*$ & \;\;8.50 \quad 7.00$^*$ & excluded \\
SS7$^{(5,6)}$\;\; & H3 & 0.25 \quad 0.50 & 1.47 \quad 1.73 & \;\;--\;\;\quad\;\;--\;\; & \;\;$1.8\times 10^{35\ast}$ & \;\;--\;\;\quad 2.22 & \;\;--\;\;\quad 2.73 & 4.00 \quad 7.00 & testable \\
SS7$^{(6,6)}$\;\; & H1 & 0.25 \quad 0.60 & \;\;1.49 \quad 1.73$^*$ & \;\;--\;\;\quad\;\;--\;\; & $2.0\times 10^{35\ast}$ & \;\;\;\;--\;\;\quad 2.22$^*$ & \;\;\;\;--\;\;\quad 2.73$^*$ & \;\;4.00 \quad 7.00$^*$ & excluded \\
SS7$^{(6,7)}$\;\; & H3 & 0.25 \quad 0.70 & 1.49 \quad 1.78 & \;\;--\;\;\quad\;\;--\;\; & $1.3\times 10^{36}$ & \;\;--\;\;\quad 2.40 & \;\;--\;\;\quad 3.00 & 4.50 \quad 7.00 & -- \\
SS7$^{(7,7)}$\;\; & H1 & 0.25 \quad 0.70 & 1.51 \quad 1.78 & \;\;--\;\;\quad\;\;--\;\; & $3.4\times 10^{36}$ & \;\;--\;\;\quad 2.40 & \;\;--\;\;\quad 3.00 & \;\;4.50 \quad 7.00$^*$ & testable \\
SS7$^{(7,8)}$\;\; & H2 & 0.25 \quad 0.80 & 1.51$^*$ \; 1.82 & \;\;--\;\;\quad\;\;--\;\; & $5.5\times 10^{36}$ & \;\;--\;\;\quad 2.56 & \;\;--\;\;\quad 3.23 & 5.50$^*$ \; 8.00 & excluded \\
SS7$^{(8,7)}$\;\; & H1 & \;\;0.25 \quad 0.80$^*$ & 1.57 \quad 1.78 & \;\;--\;\;\quad\;\;--\;\; & $1.2\times 10^{36}$ & \;\;--\;\;\quad 2.40 & \;\;--\;\;\quad 3.00 & 5.50 \quad 8.00 & excluded \\
SS7$^{(8,8)}$\;\; & H0 & 0.25 \quad 0.90 & 1.57 \quad 1.82 & \;\;--\;\;\quad\;\;--\;\; & $2.1\times 10^{37}$ & \;\;--\;\;\quad 2.56 & \;\;--\;\;\quad 3.23 & 5.50 \quad 9.00 & -- \\
SS7$^{(8,9)}$\;\; & H2 & 0.25 \quad 1.00 & 1.57$^*$\quad 1.85 & \;\;--\;\;\quad\;\;--\;\; & $2.2\times 10^{37}$ & \;\;--\;\;\quad 2.72 & \;\;--\;\;\quad 3.47 & 5.50$^*$ \; 9.00 & excluded \\
SS7$^{(9,8)}$\;\; & H1 & 0.30 \quad 0.90 & \;\;1.62 \quad 1.82$^*$ & \;\;--\;\;\quad\;\;--\;\; & $4.0\times 10^{37}$ & \;\;--\;\;\quad 2.56 & \;\;--\;\;\quad 3.23 & \;\;6.00 \quad 8.50$^*$ & excluded \\
SS7$^{(9,9)}$\;\; & H0 & 0.30 \quad 1.00 & 1.62 \quad 1.85 & \;\;--\;\;\quad\;\;--\;\; & $1.2\times 10^{38}$ & \;\;--\;\;\quad 2.72 & \;\;--\;\;\quad 3.47 & 6.00 \quad 8.50 & -- \\
SS7$^{(10,9)}$\; & H1 & 0.40 \quad 1.00 & 1.67 \quad  1.85 & \;\;--\;\;\quad\;\;--\;\; & $5.1\times 10^{38}$ & \;\;--\;\;\quad 2.72 & \;\;\;\;--\;\;\quad 3.47$^*$ & \;\;6.00 \quad 8.50$^*$ & testable \\
SS7$^{(10,10)}$ & H2 & 0.40 \quad 1.10 & 1.67 \quad  1.87 & \;\;--\;\;\quad\;\;--\;\; & $8.4\times 10^{38}$ & \;\;--\;\;\quad 2.86 & \;\;--\;\;\quad 3.66 & 6.00$^*$ \; 8.50 & testable \\
SS7$^{(11,9)}$\;\; & H1 & \;\;0.50 \quad 1.00$^*$ & 1.70 \quad 1.85 & \;\;--\;\;\quad\;\;--\;\; & $2.8\times 10^{38}$ & \;\;--\;\;\quad 2.72 & \;\;--\;\;\quad 3.47 & 6.00 \quad 8.50 & excluded \\
SS7$^{(11,10)}$ & H1 & 0.50 \quad 1.10 & 1.70 \quad 1.87 & \;\;--\;\;\quad\;\;--\;\; & $2.3\times 10^{39}$ & \;\;--\;\;\quad 2.86 & \;\;--\;\;\quad 3.66 & 6.00$^*$ \; 8.50 & testable \\
SS7$^{(11,11)}$ & H2 & 0.50 \quad 1.20 & 1.70$^*$ \; 1.89 & \;\;--\;\;\quad\;\;--\;\; & $4.6\times 10^{39}$ & \;\;--\;\;\quad 3.00 & \;\;--\;\;\quad 3.82 & 6.00$^*$ \; 8.50 & excluded \\
SS7$^{(12,10)}$ & H1 & 0.60 \quad 1.10 & \;\;1.73 \quad 1.87$^*$ & \;\;--\;\;\quad\;\;--\;\; & $6.1\times 10^{39}$ & \;\;\;\;--\;\;\quad 2.86$^*$ & \;\;\;\;--\;\;\quad 3.66$^*$ & \;\;6.00 \quad 8.50$^*$ & excluded \\
SS7$^{(12,11)}$ & H0 & 0.60 \quad 1.20 & 1.73 \quad  1.89 & \;\;--\;\;\quad\;\;--\;\; & $1.8\times 10^{40}$ & \;\;--\;\;\quad 3.00 & \;\;--\;\;\quad 3.82 & 6.00 \quad 8.50 & -- \\
SS7$^{(13,11)}$ & H1 & 0.70 \quad 1.20 & 1.76 \quad  1.89 & \;\;--\;\;\quad\;\;--\;\; & $5.1\times 10^{40}$ & \;\;--\;\;\quad 3.00 & \;\;\;\;--\;\;\quad 3.82$^*$ & \;\;6.00 \quad 8.50$^*$ & testable \\
SS7$^{(13,12)}$ & H2 & 0.70 \quad 1.30 & 1.76 \quad  1.91 & \;\;--\;\;\quad\;\;--\;\; & $1.6\times 10^{41}$ & \;\;--\;\;\quad 3.13 & \;\;--\;\;\quad 4.15 & 6.00$^*$ \; 8.50 & testable \\
SS7$^{(14,11)}$ & H1 & \;\;0.70 \quad 1.20$^*$ & 1.78 \quad  1.89 & \;\;--\;\;\quad\;\;--\;\; & $1.2\times 10^{41}$ & \;\;--\;\;\quad 3.00 & \;\;--\;\;\quad 3.82 & 6.00 \quad 8.50 & excluded \\
SS7$^{(14,12)}$ & H0 & 0.70 \quad 1.30 & 1.78 \quad  1.91 & \;\;--\;\;\quad\;\;--\;\; & $5.2\times 10^{41}$ & \;\;--\;\;\quad 3.13 & \;\;--\;\;\quad 4.15 & \;\;6.00 \quad 8.50$^*$ & testable \\
SS7$^{(14,13)}$ & H2 & 0.70$^*$ \; 1.40 & 1.78 \quad  1.92 & \;\;--\;\;\quad\;\;--\;\; & $1.1\times 10^{42}$ & \;\;--\;\;\quad 3.26 & \;\;--\;\;\quad 4.23 & 6.00 \quad 8.50 & excluded \\
SS7$^{(15,12)}$ & H1 & 0.80 \quad 1.30 & \;\;1.80 \quad 1.91$^*$ & \;\;--\;\;\quad\;\;--\;\; & $4.2\times 10^{42}$ & \;\;\;\;--\;\;\quad 3.13$^*$ & \;\;\;\;--\;\;\quad 4.15$^*$ & \;\;6.50 \quad 8.50$^*$ & excluded \\
SS7$^{(15,13)}$ & H2 & 0.80 \quad 1.40 & 1.80 \quad  1.92 & \;\;--\;\;\quad\;\;--\;\; & $6.1\times 10^{42}$ & \;\;--\;\;\quad 3.26 & \;\;--\;\;\quad 4.23 & 6.50$^*$ \; 8.50 & testable \\
SS8$^{(9,1)}$\;\;\; & H2 & 0.40 \quad 1.00 & 1.62 \quad 1.87 & \;\;--\;\;\quad\;\;--\;\; & $4.1\times 10^{35}$ & \;\;--\;\;\quad 2.56 & \;\;--\;\;\quad 3.23 & 6.00$^*$ \; 8.50 & testable \\
SS8$^{(10,1)}$\;\; & H0 & 0.40 \quad 1.00 & 1.67 \quad 1.87 & \;\;--\;\;\quad\;\;--\;\; & $1.2\times 10^{36}$ & \;\;--\;\;\quad 2.56 & \;\;--\;\;\quad 3.23 & \;\;6.00 \quad 8.50$^*$ & testable \\
SS8$^{(11,1)}$\;\; & H1 & 0.50 \quad 1.00 & \;\;1.70 \quad 1.87$^*$ & \;\;--\;\;\quad\;\;--\;\; & $1.8\times 10^{36}$ & \;\;\;\;--\;\;\quad 2.56$^*$ & \;\;\;\;--\;\;\quad 3.23$^*$ & \;\;6.00 \quad 8.50$^*$ & excluded \\
SS8$^{(12,1)}$\;\; & H1 & \;\;0.60 \quad 1.00$^*$ & 1.73 \quad 1.87 & \;\;--\;\;\quad\;\;--\;\; & $1.2\times 10^{36}$ & \;\;--\;\;\quad 2.56 & \;\;--\;\;\quad 3.23 & 6.00 \quad 8.50 & excluded \\
SS9$^{(4,1)}$\;\;\; & H2 & 0.25 \quad 0.25 & 1.57$^*$ \; 1.63 & \;\;--\;\;\quad\;\;--\;\; & $2.9\times 10^{39}$ & 1.82$^*$\quad\;--\;\; & 1.97$^*$\quad\;--\;\; & 5.50$^*$ \; 4.50 & excluded \\
SS9$^{(5,1)}$\;\;\; & H1 & 0.40 \quad 0.25 & \;\;1.67 \quad 1.63$^*$ & \;\;--\;\;\quad\;\;--\;\; & $4.1\times 10^{40}$ & 2.03 \quad\;\;--\;\; & 2.40 \quad\;\;--\;\; & \;\;6.50 \quad 5.00$^*$ & excluded \\
SS10$^{(5,1)}$\; & H3 & 0.40 \quad 0.25 & 1.67 \quad 1.63 & \;\;--\;\;\quad\;\;--\;\; & $1.4\times 10^{39}$ & 2.03 \quad\;\;--\;\; & 2.40 \quad\;\;--\;\; & 6.50 \quad 5.00 & -- \\
SS11$^{(5,1)}$\; & H3 & 0.40 \quad 0.50 & 1.67 \quad 1.76 & \;\;--\;\;\quad\;\;--\;\; & $2.8\times 10^{36}$ & 2.03 \quad 1.29 & 2.40 \quad 0.70 & 6.50 \quad 6.50 & -- \\
SS11$^{(7,2)}$\; & H2 & 0.70 \quad 1.20 & 1.78$^*$ \; 1.94 & \;\;--\;\;\quad\;\;--\;\; & $1.9\times 10^{41}$ & 2.40$^*$ \; 1.82 & 3.00$^*$ \; 1.97 & 7.50$^*$ \; 8.50 & excluded \\
SS11$^{(8,2)}$\; & H1 & 0.80 \quad 1.20 & \;\;1.78 \quad 1.94$^*$ & \;\;--\;\;\quad\;\;--\;\; & $2.4\times 10^{42}$ & \;\;2.56 \quad 1.82$^*$ & \;\;3.23 \quad 1.97$^*$ & \;\;7.50 \quad 8.50$^*$ & excluded \\
SS12$^{(7,1)}$\; & H3 & 0.80 \quad 0.25 & 1.78 \quad 1.63 & \;\;--\;\;\quad\;\;--\;\; & $1.8\times 10^{36}$ & 2.40 \quad\;\;--\;\; & 3.00 \quad\;\;--\;\; & 7.50 \quad 5.50 & -- \\
SS12$^{(11,2)}$ & H2 & 1.20 \quad 0.70 & 1.89$^*$ \; 1.76 & \;\;--\;\;\quad\;\;--\;\; & $8.6\times 10^{40}$ & 3.00$^*$\quad\;--\;\; & 3.82$^*$\quad\;--\;\; & 8.50$^*$ \; 7.00 & excluded \\
\;SS12$^{(12,2)}$ & H0 & 1.30 \quad 0.70 & 1.91 \quad 1.76 & \;\;--\;\;\quad\;\;--\;\; & $6.9\times 10^{41}$ & 3.13 \quad\;\;--\;\; & 4.15 \quad\;\;--\;\; & 8.50$^*$ \; 7.00 & testable \\
SS13$^{(3,1)}$ & H3 & 0.25 \quad 0.70 & 1.49 \quad 1.76 & \;\;--\;\;\quad & $4.2\times 10^{41}$ & 1.58 \quad 1.49 & 1.41 \quad 1.19 & 5.00 \quad 7.50 & -- \\
SS14$^{(1,2)}$ & H2 & 0.25 \quad 0.60 & 1.38$^*$ \; 1.76 & \;\;--\;\;\quad & $8.6\times 10^{37}$ & 1.82$^*$\quad\;--\;\; & 1.97$^*$\quad\;--\;\; & 5.00$^*$ \; 7.50 & excluded \\
SS15$^{(2,3)}$ & H1 & \;\;0.25 \quad 0.90$^*$ & 1.49 \quad 1.87 & \;\;--\;\;\quad & $1.1\times 10^{37}$ & 2.56 \quad\;\;--\;\; & 3.23 \quad\;\;--\;\; & 5.00 \quad 7.70 & excluded \\
SS15$^{(2,4)}$ & H2 & 0.25 \quad 1.20 & 1.49$^*$ \; 1.94 & \;\;--\;\;\quad & $6.0\times 10^{38}$ & 2.56$^*$\quad\;--\;\; & 3.23$^*$\quad\;--\;\; & 5.00$^*$ \; 8.50 & excluded \\
SS16$^{(2,4)}$ & H2 & 0.25 \quad 1.20 & 1.49 \quad 1.94 & \;\;--\;\;\quad & $2.3\times 10^{42}$ & 2.56$^*$\quad\;--\;\; & 3.23$^*$\quad\;--\;\; & 5.00$^*$ \; 8.50 & testable \\
SS17$^{(3,3)}$ & H3 & \;\;--\;\;\quad 1.90 & \;\;--\;\;\quad 2.03 & 0.94 \quad\;\;--\;\; & $4.2\times 10^{43}$ & 1.82 \quad 2.56 & 1.97 \quad 3.23 & \;\;--\;\;\quad 9.00 & -- \\
\hline
\end{longtable}
\end{center}
\end{scriptsize}

\bibliographystyle{JHEP}
\bibliography{myref}
\end{document}